\documentclass[12pt]{amsart}
\usepackage{amssymb}
\usepackage{amsmath}
\usepackage{amsfonts}
\usepackage{setspace,hyperref}
\usepackage{graphicx}
\usepackage{color}

\numberwithin{equation}{section}
\theoremstyle{plain}

\input{tcilatex}

\begin{document}
\title[Semiparametreic Models]{Semiparametric Models with Single-Index
Nuisance Parameters}
\author[K. Song]{Kyungchul Song}
\date{July 22, 2013}
\address{Vancouver School of Economics, University of British Columbia, 997
- 1873 East Mall, Vancouver, BC, V6T 1Z1, Canada}
\email{kysong@mail.ubc.ca}

\begin{abstract}
{\footnotesize In many semiparametric models, the parameter of interest is
identified through conditional expectations, where the conditioning variable
involves a single-index that is estimated in the first step. Among the
examples are sample selection models and propensity score matching
estimators. When the first-step estimator follows cube-root asymptotics, no
method of analyzing the asymptotic variance of the second step estimator
exists in the literature. This paper provides nontrivial sufficient
conditions under which the asymptotic variance is not affected by the first
step single index estimator regardless of whether it is root-n or cube-root
consistent. The finding opens a way to simple inference procedures in these
models. Results from Monte Carlo simulations show that the procedures
perform well in finite samples.}

{\footnotesize \ }

{\footnotesize \noindent \textsc{Key words.} Sample selection model;
conditional median restrictions; matching estimators; maximum score
estimation; cube-root asymptotics; generated regressors \newline
}


\end{abstract}

\maketitle

\section{Introduction}

Many empirical studies use a number of covariates to deal with the problem
of endogeneity. Using too many covariates in nonparametric estimation,
however, tends to worsen the quality of the empirical results significantly.
A promising approach in this situation is to introduce a single-index
restriction so that one can retain flexible specification while avoiding the
curse of dimensionality. The single-index restriction has long attracted
attention in the literature.\footnote{%
For example, Klein and Spady (1993) and Ichimura (1993) proposed \textit{M}%
-estimation approaches to estimate the single-index, and Stoker (1986) and
Powell, Stock and Stoker (1989) proposed estimation based on average
derivatives. See also H\"{a}rdle, Hall, and Ichimura (1993), H\"{a}rdle and
Tsybakov (1993), Horowitz and H\"{a}rdle (1996), Fan and Li (1996) and
Hristache, Juditsky and Spokoiny (2001).}

Most literatures deal with a single-index model as an isolated object,
whereas empirical researchers often need to use the single-index
specification in the context of estimating a larger model. An example is a
structural model in labor economics that requires a prior estimation of
components such as wage equations. When single-index components are nuisance
parameters that are plugged into the second step estimation of a finite
dimensional parameter of interest, the introduction of single-index
restrictions does not improve the convergence rate of the estimated
parameter of interest which already achieves the parametric rate of $\sqrt{n}%
.$ Nevertheless, the use of a single-index restriction in such a situation
still has its own merits. After its adoption, the model requires weaker
assumptions on the nonparametric function and on the kernel function. This
merit becomes prominent when the nonparametric function is defined on a
space of a large dimension and stronger conditions on the nonparametric
function and higher-order kernels are required. (See Hristache, Juditsky and
Spokoiny (2001) for more details.)

This paper focuses on semiparametric models, where the parameter of interest
is identified through a conditional expectation function and the
conditioning variable involves a single-index with an unknown finite
dimensional nuisance parameter. We assume that there is a consistent first
step estimator of this nuisance parameter. In this situation, a natural
procedure is a two step estimation, where one estimates the single-index
first, and uses it to estimate the parameter of interest in the second step.
Among the examples are sample selection models and propensity score matching
estimators. The examples will be discussed in detail later.

A distinctive feature of the framework of this paper is that the first step
estimator of a single-index is allowed to be either $\sqrt{n}$-consistent or 
$\sqrt[3]{n}$-consistent. The latter case of $\sqrt[3]{n}$-consistent single
index estimators is particularly interesting, for the framework includes new
models that have not been studied in the literature, such as the sample
selection model with conditional median restrictions, or propensity score
matching estimators with conditional median restrictions. These conditional
median restrictions often lead to a substantial relaxation of the existing
assumptions that have been used in the literature.\footnote{%
For example, the semiparametric sample selection model in Newey, Powell and
Walker (1990) assumes that the error term in the selection equation is
independent of observed covariates. Also, parametric specifications of
propensity scores in the literature of program evaluations (such as logit or
probit specifications) assume that the error term in the program
participation equation is independent of observed covariates. (See Heckman,
Ichimura, Smith and Todd (1998) for example.) In these situations, the
assumption of the conditional median restriction is a weaker assumption
because it allows for stochastic dependence between the error term and the
observed covariates.}

Dealing with the case of a nuisance parameter that follows cube-root
asymptotics of Kim and Pollard (1990) in two step estimation is challenging.
In typical two step estimation, the asymptotic variance of the second step
estimator involves an additional term due to the first step estimation of
the single-index component (e.g. Newey and McFadden (1994).) Unless this
term is shown to be negligible, one needs to compute this additional term by
first finding the asymptotic linear representation of the first step
estimator. However, in the case of a first step estimator that follows
cube-root asymptotics, there does not exist such an asymptotic linear
representation.

The main contribution of this paper is to provide a set of conditions under
which the first step estimator, regardless of whether it is $\sqrt{n}$%
-consistent or $\sqrt[3]{n}$-consistent, does not have an impact on the
asymptotic variance of the second step estimator. This result is convenient,
because under these conditions, one can simply compute the asymptotic
variance as if one knows the true nuisance parameter in the single-index.

The result of this paper is based on a recent finding by the author (Song
(2012)) which offers generic conditions under which conditional expectation
functionals are very smooth. This smoothness is translated in our situation
into insensitivity of the parameter of interest at a local perturbation of
the single-index nuisance parameter.

To illustrate the usefulness of the result, this paper applies it to new
semiparametric models such as semiparametric sample selection models with
conditional median restrictions, and single-index matching estimators with
conditional median restrictions. This paper offers procedures to obtain
estimators and asymptotic variance formulas for the estimators.

This paper presents and discusses results from Monte Carlo simulation
studies. The main focus of these studies lies on whether the asymptotic
negligibility of the first step estimator's impact remains in force in
finite samples. For this, it is investigated whether the estimators and the
confidence sets based on the proposed asymptotic covariance matrix formula
performs reasonably well in finite samples. Simulation results demonstrate
clearly that they do so.

The main result of this paper is closely related to the literature of
so-called \textit{generated regressors} in nonparametric or semiparametric
models. For example, Newey, Powell, and Vella (1999) and Das, Newey, and
Vella (2003) considered nonparametric estimation of simultaneous equation
models. Li and Wooldridge (2002) analyzed partial linear models with
generated regressors when the estimated parameters in the generated
regressors are $\sqrt{n}$-consistent. Rilstone (1996) and Sperlich (2009)
studied nonparametric function estimators that involve generated regressors.
Recent contributions by Hahn and Ridder (2010) and Mammen, Rothe, and
Schienle (2012) offer a general analysis of the issue with generated
regressors in nonparametric or semiparametric models. None of these papers
considered generated regressors with coefficient estimators that follow
cube-root asymptotics.

The paper is organized as follows. The paper defines the scope, introduces
examples, and explains the main idea of this paper in the next section. Then
Section 3 presents the formal result of the asymptotic distribution theory,
and discusses their implications for exemplar models. Section 4 discusses
Monte Carlo simulation results, and Section 5 presents an empirical
illustration based on a simple female labor supply model. Some technical
proofs are found in the Appendix.

\section{The Scope, Examples, and the Main Idea}

\subsection{The Scope of the Paper}

Let us define the scope of the paper. Suppose that $W\equiv (W_{1},\cdot
\cdot \cdot ,W_{L})^{\top }\in \mathbf{R}^{L},$ $S$ is a $d_{S}\times
d_{\varphi }$ random matrix, and $X\in \mathbf{R}^{d}$ is a random vector,
where all three random quantities $W$, $S,$ and $X$, are assumed to be
observable. We let $X=[X_{1}^{\top },X_{2}^{\top }]^{\top }\in \mathbf{R}%
^{d_{1}+d_{2}}$, where $X_{1}$ is a continuous random vector and $X_{2}$ is
a discrete random vector taking values from $\{x_{1},\cdot \cdot \cdot
,x_{M}\}$. Let $\Theta \subset \mathbf{R}^{d}$ be the space of a nuisance
parameter $\theta _{0}$ that is known to be identified. Denote $U_{\theta
}\equiv F_{\theta }(X^{\top }\theta )$, where $F_{\theta }$ is the CDF of $%
X^{\top }\theta $. We assume that $X^{\top }\theta $ is a continuous random
variable for all $\theta $ in a neighborhood of $\theta _{0}$. Given an
observed binary variable $D\in \{0,1\}$, we define%
\begin{equation}
\mu _{\theta }(U_{\theta })\equiv \mathbf{E}\left[ W|U_{\theta },D=1\right] ,
\label{mu}
\end{equation}%
and when $\theta =\theta _{0}$, we simply write $\mu _{0}(U_{0})$, where $%
U_{0}\equiv F_{\theta _{0}}(X^{\top }\theta _{0})$. The support of a random
vector is defined to be the smallest closed set in which the random vector
takes values with probability one. For $m=1,\cdot \cdot \cdot ,M$, let $%
\mathcal{S}_{m}$ be the support of $X1\{X_{2}=x_{m},D=1\}$. and $\mathcal{S}%
_{W}$ be the support of $W$, and let $\varphi :\mathcal{S}_{W}\rightarrow 
\mathbf{R}^{d_{\varphi }}$ be a known map that is twice continuously
differentiable with bounded derivatives on the interior of the support of $%
\mathbf{E}[W|X,D=1]$. Then we define a map $a:\Theta \rightarrow \mathbf{R}%
^{d_{S}}$ by%
\begin{equation}
a(\theta )\equiv \mathbf{E}\left[ S\cdot \varphi (\mu _{\theta }(U_{\theta
}))|D=1\right] ,\ \theta \in \Theta \text{.}  \label{gamma}
\end{equation}%
The general formulation admits the case without conditioning on $D=1$ in
which case it suffices to put $D=1$ everywhere.\footnote{%
The conditions for the identification of $\theta _{0}$ in many examples of
semiparametric models is already established. See Horowitz (2009). The
identification of $\theta _{0}$ in this paper's context arises often in
binary choice models. See Chapter 4 of Horowitz (2009) for the
identification analysis for the binary choice models.}

This paper focuses on semiparametric models where the parameter of interest,
denoted by $\beta _{0},$ is identified as follows:%
\begin{equation}
\beta _{0}=H(a(\theta _{0}),b_{0}),  \label{param}
\end{equation}%
where $H:\mathbf{R}^{d_{S}}\times \mathbf{R}^{d_{b}}\rightarrow \mathbf{R}%
^{d_{\beta }}$ is a map that is fully known, continuously differentiable in
the first argument, and $b_{0}$ is a $d_{b}$ dimensional parameter that does
not depend on $\theta _{0}$ and is consistently estimable. We will see
examples of $\beta _{0}$ shortly.

Throughout this paper, we assume that there is an estimator $\hat{\theta}$
for $\theta _{0}$ which is either $\sqrt{n}$-consistent or $\sqrt[3]{n}$%
-consistent. A natural estimator of $\beta _{0}$ is obtained by 
\begin{equation*}
\hat{\beta}\equiv H(\hat{a}(\hat{\theta}),\hat{b}),
\end{equation*}%
where $\hat{a}(\theta )$ is an estimator of $a(\theta )$ and $\hat{b}$ is a
consistent estimator of $b_{0}$. The estimator $\hat{a}(\theta )$ can be
obtained by using nonparametric estimation of conditional expectation $%
\mathbf{E}\left[ W|U_{\theta },D=1\right] $. For future reference, we denote 
\begin{equation*}
\tilde{\beta}\equiv H(\hat{a}(\theta _{0}),\hat{b}),
\end{equation*}%
an infeasible estimator using $\theta _{0}$ in place of $\hat{\theta}$. When 
$\hat{\theta}$ is $\sqrt[3]{n}$-consistent, it is not clear whether $\sqrt{n}%
(\hat{\beta}-\beta )$ will be asymptotically normal. In fact, it is not even
clear whether $\hat{\beta}$ will be $\sqrt{n}$-consistent.

The main contribution of this paper is to provide conditions under which,
whenever $\hat{\theta}=\theta _{0}+O_{P}(n^{-1/3})$ and%
\begin{equation}
\sqrt{n}(\tilde{\beta}-\beta _{0})\overset{d}{\rightarrow }N(0,V),  \label{d}
\end{equation}%
it follows that%
\begin{equation*}
\sqrt{n}(\hat{\beta}-\beta _{0})\overset{d}{\rightarrow }N(0,V).
\end{equation*}%
This result is very convenient, because the computation of the asymptotic
variance matrix $V$ in (\ref{d}) can be done, following the standard
procedure.

\subsection{Examples}

\subsubsection{Example 1: Sample Selection Models with Conditional Median
Restrictions}

Let $Y^{\ast }$ be an outcome variable which is related to $Z\in \mathbf{R}%
^{d_{Z}},$ a vector of covariates, as follows: 
\begin{equation*}
Y^{\ast }=Z^{\top }\beta _{0}+v\text{,}
\end{equation*}%
where $v$ denotes an unobserved factor that affects the outcome. The
econometrician observes $Y^{\ast }$ only when a selection indicator $D\in
\{0,1\}$ assumes number one, so that as for observed outcome $Y$, we write%
\begin{equation*}
Y=Y^{\ast }\cdot D.
\end{equation*}%
This paper specifies $D$ as follows:%
\begin{equation}
D=1\{X^{\top }\theta _{0}>\varepsilon \},  \label{se}
\end{equation}%
where $\varepsilon $ is an unobserved component, and $\theta _{0}\in \mathbf{%
R}^{d}$ an unknown parameter.

Sample selection models and their inference procedures have been extensively
studied in the literature. The early generation of these models impose
parametric distributional assumptions on the unobserved components (Heckman
(1974)). Gallant and Nychka (1987), Cosslett (1990) and many others (e.g.
Newey, Powell, and Walker (1990), and Das, Newey and Vella (1999)) analyzed
semiparametric or nonparametric models that do not require parametric
distributional assumptions. A common feature for these various models is the
following assumption:\footnote{%
An exception to this assumption is Chen and Khan (2003) who considered
semiparametric sample selection models with conditional heteroskedasticity,
and proposed a three-step estimation procedure.}%
\begin{equation}
\varepsilon \ \text{is\ independent\ of\ }X.  \label{CA}
\end{equation}%
Condition (\ref{CA}) is mainly used to identify $\theta _{0}$ through a
single-index restriction $\mathbf{E}[D|X]=\mathbf{E}[D|X^{\top }\theta _{0}]$
or a parametric restriction $\mathbf{E}[D|X]=F(X^{\top }\theta _{0})$ for
some known CDF $F$. Define for $\theta $ in a neighborhood of $\theta _{0},$%
\begin{eqnarray*}
S_{ZZ}(\theta ) &\equiv &\mathbf{E}\left[ ZZ^{\top }|D=1\right] -\mathbf{E}%
\left[ Z\cdot \mathbf{E}[Z^{\top }|U_{\theta },D=1]|D=1\right] \text{ and} \\
S_{ZY}(\theta ) &\equiv &\mathbf{E}[ZY|D=1]-\mathbf{E}\left[ Z\cdot \mathbf{E%
}[Y|U_{\theta },D=1]|D=1\right] .
\end{eqnarray*}%
We consider the following assumptions.\bigskip

\noindent \textsc{Assumption SS0} \textsc{:}\textbf{\ }(i) $(\varepsilon
,v)\ $is conditionally independent of $Z$ given $X^{\top }\theta _{0}$.

\noindent (ii) $Med(\varepsilon |X)=0,$ a.e., where $Med(\varepsilon |X)$
denotes the conditional median of $\varepsilon $ given $X$.

\noindent (iii) The smallest eigenvalue of $S_{ZZ}(\theta )$ is bounded away
from zero uniformly over $\theta $ in a neighborhood of $\theta _{0}.$%
\bigskip

Assumption SS0(i) is a form of an index exogeneity condition. Such an
assumption has been used in various forms in the literature (e.g. Powell
(1994).) The distinctive feature of this model stems from Assumption SS0(ii)
which substantially relaxes Condition (\ref{CA}). The relaxation allows the
individual components of $X$ and $\varepsilon $ to be stochastically
dependent. Assumption SS0(iii) is slightly stronger than the usual condition
that $S_{ZZ}(\theta _{0})$ is invertible.

Under Assumptions SS0(i) and (iii), we can write the equation for observed
outcomes as a partial linear regression model, and follow Robinson (1988) to
identify $\beta _{0}$ as%
\begin{equation*}
\beta _{0}=S_{ZZ}^{-1}(\theta _{0})\cdot S_{ZY}(\theta _{0}),
\end{equation*}%
once $\theta _{0}$ is identified. To see that this $\beta _{0}$ is a special
case of (\ref{param}), let $\mathbf{1}_{2}$ be a $2\times 1$ vector of ones, 
$S=\mathbf{1}_{2}\otimes Z$, (the notation $\otimes $ represents the
Kronecker product of matrices) and define $W=[Y;Z^{\top }]$, $\mu (U_{\theta
})=\mathbf{E}[W|U_{\theta },D=1]$, 
\begin{eqnarray*}
a(\theta ) &=&\mathbf{E}\left[ S\cdot \varphi (\mu (U_{\theta }))|D=1\right]
,\text{ and} \\
b_{0} &=&\mathbf{E}\left[ S\cdot W|D=1\right] ,
\end{eqnarray*}%
where $\varphi :\mathbf{R}^{d_{Z}+1}\rightarrow \mathbf{R}^{d_{Z}+1}$ is an
identity map. Note that $a(\theta )$ and $b_{0}$ are $2d_{Z}\times (d_{Z}+1)$
matrices. Furthermore, $b_{0}$ does not depend on $\theta $. Given $%
2d_{Z}\times (d_{Z}+1)$ matrices $a$ and $b$, we denote $a_{22}$ and $b_{22}$
to be the $d_{Z}\times d_{Z}$ lower-right sub-blocks of $a$ and $b,$ and
denote $a_{11}$ and $b_{11}$ to be the $d_{Z}\times 1$ upper-left sub-blocks
of $a$ and $b$. Then define%
\begin{equation*}
H(a,b)=(b_{22}-a_{22})^{-1}(b_{11}-a_{11}),
\end{equation*}%
whenever $b_{22}-a_{22}$ is invertible. We can reformulate the
identification result as follows:%
\begin{equation*}
\beta _{0}=H(a(\theta _{0}),b_{0}),
\end{equation*}%
which shows that $\beta _{0}$ is a special case of (\ref{param}).

\subsubsection{Example 2: Single-Index Matching Estimators of Treatment
Effects on the Treated}

Among various estimators of treatment effects used in the studies on program
evaluations, matching estimators have been widely studied and used. (See
Dehejia and Wahba (1998) and Heckman, Ichimura and Todd (1997, 1998) and
references therein for matching methods in general.) While many studies of
econometric methodologies use nonparametric specification of the propensity
score (e.g. Hahn (1998), Hirano, Imbens and Ridder (2003)), a single-index
restriction on the propensity score can be useful in avoiding curse of
dimensionality.

When the propensity score is specified by logit or probit assumptions, the
propensity score is strictly increasing in the single-index. In general,
when the propensity score satisfies a single-index restriction and is a
strictly increasing function of the single-index, identification of the
average treatment effects on the treated through propensity score matching
is equivalent to the identification through single-index matching, because
the $\sigma $-field generated by the propensity score is the same as that
generated by the single-index.

In the current example, we develop what this paper calls a \textit{%
single-index matching estimator}. The main merit of the single-index
matching estimators is that the estimator does not require a parametric
distributional assumption on the propensity score, while avoiding the curse
of dimensionality. The distinctive feature of the estimator as a result of
this paper's framework is that the single-index component is allowed to be
estimable only at the cube-root rate. Such a case is relevant when the
assumption of independence between the observed component and the unobserved
component in the propensity score is relaxed into the assumption of
conditional median independence.

Let $Y_{1}$ and $Y_{0}$ be potential outcomes of treated and untreated
individuals and $Z\in \{0,1\}$ the treatment status, where $Z=1$ for the
status of treatment and $Z=0$ for the status of non-treatment.\footnote{%
The common notation for the treatment status is $D$, but the treatment
status does not play the same role as $D$ in (\ref{mu}). Hence we choose a
different notation, $Z$, here.} The econometrician observes $Z$ and $%
Y=Y_{1}Z+Y_{0}(1-Z)$. The parameter of interest is $\beta _{0}=\mathbf{E}%
[Y_{1}-Y_{0}|Z=1],$ i.e., the treatment effect on the treated. We assume
that $X$ is a vector of observed covariates in $\mathbf{R}^{d}$ and%
\begin{equation*}
Z=1\left\{ X^{\top }\theta _{0}\geq \varepsilon \right\} ,
\end{equation*}%
where $\varepsilon $ denotes the unobserved factor that affects the
treatment status, and $\theta _{0}$ is an unknown parameter. Define $%
U_{\theta }=F_{\theta }(X^{\top }\theta )$, where $F_{\theta }$ is the CDF
of $X^{\top }\theta $ and is assumed to be strictly increasing, and we write
simply $U_{0}=U_{\theta _{0}}$. We also define the propensity score $%
P(U_{\theta })=P\{Z=1|U_{\theta }\}$.\bigskip 

\noindent \textsc{Assumption SM0} \textsc{:}\textbf{\ }(i) $\mathbf{E}%
[Y_{0}|U_{0},Z=0]=\mathbf{E}[Y_{0}|U_{0},Z=1].$

\noindent (ii) There exists $\eta >0$ such that $\eta \leq P(U_{0})\leq
1-\eta .$

\noindent (iii) $Med(\varepsilon |X)=0$.\bigskip

The first condition in Assumption SM0(i) is weaker than the unconfoundedness
assumption, i.e., the assumption of conditional independence between $%
(Y_{1},Y_{0})$ and $Z$ given $U_{0}$, and as noted by Heckman, Ichimura, and
Todd (1997), this assumption together with Assumption SM0(ii) suffices for
identification of the average treatment effect on the treated. Assumption
SM0(ii) requires that the propensity score is away from 0 and 1. The new
feature of the model is Assumption SM0(iii) which says that the conditional
median of the observed component in the propensity score is zero once $X$ is
conditioned on. This condition is much weaker than the common assumption
that $\varepsilon $ and $X$ are independent.

Under Assumption SM0, we can identify $\beta _{0}$ as follows:%
\begin{equation}
\beta _{0}=\mathbf{E}\left[ Y-\mathbf{E}[Y|U_{0},Z=0]|Z=1\right] .
\label{beta0}
\end{equation}%
It is not immediately seen that $\beta _{0}$ can be written in the form of (%
\ref{param}), because the conditioning on $Z=0$ in the inner conditional
expectation is different from the conditioning on $Z=1$ in the outer
conditional expectation. To write it in the form of (\ref{param}), rewrite $%
\beta _{0}$ as 
\begin{equation*}
\frac{\mathbf{E}\left[ YZ\right] }{P\{Z=1\}}-\frac{1}{P\{Z=1\}}\cdot \mathbf{%
E}\left[ Z\cdot \frac{\mathbf{E}[Y(1-Z)|U_{0}]}{1-P(U_{0})}\right] .
\end{equation*}%
Define $W=[Y(1-Z),Z]^{\top }$, and write $\mu _{\theta }(U_{\theta })=%
\mathbf{E}[W|U_{\theta }]$. Let%
\begin{equation*}
b_{0}=\left[ 
\begin{array}{c}
\mathbf{E}\left[ YZ\right] /P\{Z=1\} \\ 
1/P\{Z=1\}%
\end{array}%
\right] \text{, and }a(\theta )=\mathbf{E}\left[ Z\varphi \left( \mu
_{\theta }(U_{\theta })\right) \right] ,
\end{equation*}%
where $\varphi :\mathbf{R}\times (0,1)\rightarrow \mathbf{R}$ is defined to
be $\varphi (x,z)=x/(1-z)$ for $(x,z)\in \mathbf{R}\times (0,1)$. Note that $%
b_{0}$ does not depend on $\theta .$ Let $H:\mathbf{R}\times \mathbf{R}%
^{2}\rightarrow \mathbf{R}$ be defined as $H(a,b)=b_{1}-b_{2}\cdot a$, where 
$a\in \mathbf{R}$ and $b=[b_{1},b_{2}]^{\top }\in \mathbf{R}^{2}$. Then, we
can write $\beta _{0}=H\left( a(\theta _{0}),b_{0}\right) ,$ i.e., in the
form of (\ref{param}) with $D$ there simply replaced by $1$.

\subsection{A Heuristic Summary of the Main Idea}

As previously mentioned, the main contribution of this paper is to provide
nontrivial sufficient conditions under which the first step estimator error
of $\hat{\theta}$ does not affect the asymptotic distribution of $\hat{\beta}
$, regardless of whether $\hat{\theta}$ is $\sqrt{n}$-consistent or $\sqrt[3]%
{n}$-consistent. The development is based on the finding due to Song (2012)
that under regularity conditions that are to be made precise later, the
function $a(\theta )$ defined in (\ref{gamma}) is very smooth in $\theta $
in a neighborhood of $\theta _{0}$.\footnote{%
Song (2012) considered explicitly the case where $X$ has both discrete and
continuous components. Certainly, the result holds when $X$ has only
continuous components, because the discrete component can be replaced by a
constant. However, the result requires that $X$ have at least one continuous
component.} More specifically, under regularity conditions, there exist $C>0$
and $\varepsilon \in (0,1/2]$ such that for each $\eta \in (0,\varepsilon ],$%
\begin{equation}
\sup_{\theta \in B(\theta _{0};\eta )}\left\Vert a(\theta )-a(\theta
_{0})\right\Vert \leq C\eta ^{2},  \label{bd6}
\end{equation}%
where $B(\theta _{0};\eta )$ denotes the $\eta $-ball around $\theta _{0}$,
i.e., $B(\theta _{0};\eta )\equiv \{\theta \in \Theta :||\theta -\theta
_{0}||<\eta \}$. The novel feature of the above bound lies in the fact that
the exponent of $\eta $ is $2$ (not 1), which says that the map $a$ is 
\textit{very smooth} in a neighborhood of $\theta _{0}$.\footnote{%
The inequality in (\ref{bd6})$\ $is obtained without assuming
differentiability of $\mu _{\theta }(u)=\mathbf{E}\left[ W|U_{\theta }=u%
\right] $ in $\theta \in \Theta $. The low level conditions for the bound in
(\ref{bd6}) are provided in Lemma A1 in Appendix.}

To see how this result serves our purpose, we write%
\begin{eqnarray*}
||\hat{\beta}-\tilde{\beta}|| &=&||H(\hat{a}(\hat{\theta}),\hat{b})-H(\hat{a}%
(\theta _{0}),\hat{b})|| \\
&\leq &C||\hat{a}(\hat{\theta})-\hat{a}(\theta _{0})||+o_{P}(||\hat{a}(\hat{%
\theta})-\hat{a}(\theta _{0})||),
\end{eqnarray*}%
by the continuous differentiability of map $H$. As for the last term,
observe that%
\begin{eqnarray*}
\hat{a}(\hat{\theta})-\hat{a}(\theta _{0}) &=&\hat{a}(\hat{\theta})-a(\hat{%
\theta})-\{\hat{a}(\theta _{0})-a(\theta _{0})\} \\
&&+a(\hat{\theta})-a(\theta _{0}) \\
&\equiv &A_{n}+B_{n},\text{ say.}
\end{eqnarray*}%
As long as $||\hat{\theta}-\theta _{0}||=o_{P}(1)$, the term $A_{n}$ can be
shown to be $o_{P}(1/\sqrt{n})$ using the standard arguments of stochastic
equicontinuity.\footnote{%
See the derivation (3.8) in Andrews (1994) for example.} For this, standard
arguments use empirical process theory. (e.g. Andrews (1994).) To use
empirical process theory, the crucial step is to obtain the asymptotic
linear representation of $\sqrt{n}\{\hat{a}(\theta )-a(\theta )\}$, that is, 
\begin{equation}
\sqrt{n}\{\hat{a}(\theta )-a(\theta )\}=\frac{1}{\sqrt{n}}\sum_{i=1}^{n}\psi
_{i}(\theta )+o_{P}(1)  \label{ex1}
\end{equation}%
for some $\psi _{i}(\theta )$ such that $\mathbf{E}\psi _{i}(\theta )=0$ and 
$\psi _{i}(\theta )$'s are i.i.d. across $i$'s, where $o_{P}(1)$ is uniform
local around $\theta _{0}$, and apply the maximal inequality to obtain the
stochastic equicontinuity of the empirical process (i.e. the sum on the
right hand side.) To see how this works intuitively, note that 
\begin{equation*}
\sqrt{n}\left( \{\hat{a}(\theta )-a(\theta )\}-\{\hat{a}(\theta
_{0})-a(\theta _{0})\}\right) =\frac{1}{\sqrt{n}}\sum_{i=1}^{n}\left[ \psi
_{i}(\theta )-\psi _{i}(\theta _{0})\right] +o_{P}(1).
\end{equation*}%
Hence the asymptotic variance of the left hand side is equal to 
\begin{equation*}
\mathbf{E}\left( \psi _{i}(\theta )-\psi _{i}(\theta _{0})\right) ^{2}.
\end{equation*}%
As $\theta \rightarrow \theta _{0}$, the variance goes to zero, under minor
regularity conditions for $\psi _{i}(\theta )$. Hence we obtain%
\begin{equation*}
\{\hat{a}(\theta )-a(\theta )\}-\{\hat{a}(\theta _{0})-a(\theta
_{0})\}=o_{P}(n^{-1/2})
\end{equation*}%
as $\theta \rightarrow \theta _{0}$. The actual argument is more complex
than this, because what we have here is $\hat{\theta}\rightarrow _{P}\theta
_{0}$, not $\theta \rightarrow \theta _{0}$. For this, standard arguments
use empirical process theory to fill this gap. See Andrews (1994) for more
details. The crucial step here is the derivation of the asymptotic linear
representation in (\ref{ex1}). The asymptotic linear representation based on
a symmetrized nearest neighborhood estimator that is proved in the Appendix.

As for $B_{n}$, the result in (\ref{bd6})\ implies that with probability
approaching one,%
\begin{equation}
||a(\hat{\theta})-a(\theta _{0})||\leq \sup_{\theta \in \Theta :||\theta
-\theta _{0}||\leq \eta _{n}}\left\Vert a(\theta )-a(\theta _{0})\right\Vert
\leq C\eta _{n}^{2},\text{ for some }C>0,  \label{bd8}
\end{equation}%
if $||\hat{\theta}-\theta _{0}||\leq \eta _{n}$ with probability approaching
one. If $\hat{\theta}=\theta _{0}+O_{P}(n^{-1/3})$, we find that by taking $%
\eta _{n}=n^{-1/3}\log n$, the left-end term of (\ref{bd8}) is $o_{P}(1/%
\sqrt{n})$. Therefore, we conclude that%
\begin{equation*}
||\hat{\beta}-\tilde{\beta}||=o_{P}(1/\sqrt{n}).
\end{equation*}%
This implies that if$\ \sqrt{n}(\tilde{\beta}-\beta _{0})$ has an asymptotic
normal distribution, the quantity $\sqrt{n}(\hat{\beta}-\beta _{0})$ has the
same asymptotic normal distribution.

Let us now consider (\ref{bd6}). Since $a(\theta )$ depends on $\theta $
only through the $\sigma $-field generated by $X^{\prime }\theta $, proving (%
\ref{bd6}) is far from obvious. The sketch of the proof in Song (2012) is as
follows. First, we consider a discretized version of $U_{\theta }$ and $%
U_{0} $, say, $U_{\theta }^{\Delta }$ and $U_{0}^{\Delta }$, where $\Delta
>0 $ represents the size of steps in the grid points which $U_{\theta
}^{\Delta }$ and $U_{0}^{\Delta }$ take values from. For simplicity, we let $%
d_{\varphi }=L=1$, $\varphi $ be an identity map, and $D=1$. Then for $%
\theta $ in a neighborhood of $\theta _{0}$, $\left\vert a(\theta )-a(\theta
_{0})\right\vert $ is bounded by%
\begin{eqnarray*}
&&\left\vert \mathbf{E}\left[ S\cdot \{\mathbf{E}[W|U_{\theta }]-\mathbf{E}%
[W|U_{\theta }^{\Delta }]\}\right] \right\vert +\left\vert \mathbf{E}\left[
S\cdot \{\mathbf{E}[W|U_{\theta }^{\Delta }]-\mathbf{E}[W|U_{0}^{\Delta }]\}%
\right] \right\vert \\
&&+\left\vert \mathbf{E}\left[ S\cdot \{\mathbf{E}[W|U_{0}^{\Delta }]-%
\mathbf{E}[W|U_{0}]\}\right] \right\vert \\
&=&|A_{1n}|+|A_{2n}|+|A_{3n}|\text{, say.}
\end{eqnarray*}%
Since $\sigma (U_{\theta }^{\Delta })\subset \sigma (U_{\theta })$ (where $%
\sigma (U_{\theta }^{\Delta })$ and $\sigma (U_{\theta })$ denoting the $%
\sigma $-fields of $U_{\theta }^{\Delta }$ and $U_{\theta }$), we use the
law of iterated conditional expectations to write $A_{1n}$ as%
\begin{eqnarray*}
&&\mathbf{E}\left[ \mathbf{E}[S|U_{\theta }]\cdot \{\mathbf{E}[W|U_{\theta
}]-\mathbf{E}[W|U_{\theta }^{\Delta }]\}\right] \\
&=&\mathbf{E}\left[ \left( \mathbf{E}[S|U_{\theta }]-\mathbf{E}[S|U_{\theta
}^{\Delta }]\right) \cdot \{\mathbf{E}[W|U_{\theta }]-\mathbf{E}[W|U_{\theta
}^{\Delta }]\}\right] .
\end{eqnarray*}%
Therefore, by Cauchy-Schwarz inequality, we find that%
\begin{equation}
|A_{1n}|\leq \sqrt{\mathbf{E}\left[ \left( \mathbf{E}[S|U_{\theta }]-\mathbf{%
E}[S|U_{\theta }^{\Delta }]\right) ^{2}\right] }\sqrt{\mathbf{E}\left[
\left( \mathbf{E}[W|U_{\theta }]-\mathbf{E}[W|U_{\theta }^{\Delta }]\right)
^{2}\right] }.  \label{CS}
\end{equation}%
Note that as $\Delta $ becomes smaller $\mathbf{E}[S|U_{\theta }]$ and $%
\mathbf{E}[S|U_{\theta }^{\Delta }]$ become closer. In fact, under
regularity conditions, it can be shown that%
\begin{eqnarray*}
\sqrt{\mathbf{E}\left[ \left( \mathbf{E}[W|U_{\theta }]-\mathbf{E}%
[W|U_{\theta }^{\Delta }]\right) ^{2}\right] } &\leq &C\Delta \text{ and} \\
\sqrt{\mathbf{E}\left[ \left( \mathbf{E}[S|U_{\theta }]-\mathbf{E}%
[S|U_{\theta }^{\Delta }]\right) ^{2}\right] } &\leq &C\Delta ,
\end{eqnarray*}%
for some $C>0$. (See Lemma 3 of Song (2012).) Hence from (\ref{CS}), we
conclude that%
\begin{equation*}
|A_{1n}|\leq C_{1}\Delta ^{2}\text{,}
\end{equation*}%
for some $C_{1}>0$. Applying the same arguments to $A_{3n}$, we also obtain
that%
\begin{equation*}
|A_{3n}|\leq C_{2}\Delta ^{2}\text{,}
\end{equation*}%
for some $C_{2}>0$.

Now as for $A_{2n}$, one can expect that $\mathbf{E}[W|U_{\theta }^{\Delta }]
$ and $\mathbf{E}[W|U_{0}^{\Delta }]$ will become close to each other as $%
\theta $ becomes closer to $\theta _{0}$. How fast they become closer to
each other will also depend on how fine the discretization is. In fact,
using Lemma 4 of Song (2012), one can show that for some $C_{3}>0,$%
\begin{equation*}
|A_{2n}|\leq C_{3}\Delta ||\theta -\theta _{0}||,
\end{equation*}%
whenever $||\theta -\theta _{0}||\leq C_{4}\Delta $ for some constant $%
C_{4}>0$. Hence $\mathbf{E}[W|U_{\theta }^{\Delta }]$ and $\mathbf{E}%
[W|U_{0}^{\Delta }]$ become closer to each other faster when $\Delta $ is
smaller. (Note that the bound requires that $||\theta -\theta _{0}||\leq
C_{4}\Delta $, and hence one cannot make $\Delta $ smaller without making $%
\theta $ closer to $\theta _{0}$.) Thus, we collect the results to obtain
the following bound:%
\begin{equation}
\left\vert a(\theta )-a(\theta _{0})\right\vert \leq C_{5}\Delta
^{2}+C_{6}\Delta ||\theta -\theta _{0}||,  \label{bd}
\end{equation}%
for some constants $C_{5}>0$ and $C_{6}>0.$ Since one can take the
discretization step $\Delta $ arbitrarily small as long as the condition $%
||\theta -\theta _{0}||\leq C_{4}\Delta $ is satisfied, we take $\Delta
=C_{4}^{-1}||\theta -\theta _{4}||$ to obtain the last bound in (\ref{bd}) as%
\begin{equation*}
C_{5}C_{4}^{-2}||\theta -\theta _{0}||^{2}+C_{6}C_{4}^{-1}||\theta -\theta
_{0}||^{2}\leq C||\theta -\theta _{0}||^{2},
\end{equation*}%
where $C=C_{5}C_{4}^{-2}+C_{6}C_{4}^{-1}$. Thus we obtain the desired result%
\begin{equation*}
\left\vert a(\theta )-a(\theta _{0})\right\vert \leq C||\theta -\theta
_{0}||^{2},
\end{equation*}%
for some $C>0$, and hence (\ref{bd6}).\footnote{%
The heuristics here uses the assumption that $\varphi $ is a linear map.
When $\varphi $ is nonlinear yet twice continuously differentiable, we can
linearize it to obtain a similar bound. (See Song (2012) for details.)}

\section{Inference and Asymptotic Theory}

\subsection{General Asymptotic Theory}

Let us consider an estimation method of $\hat{\beta}$. Instead of putting
forth high level conditions, we choose a specific estimator $\hat{\beta}$
and provide low level conditions. Suppose that we are given an estimator $%
\hat{\theta}$ of $\theta _{0}$ such that $\hat{\theta}=\theta
_{0}+O_{P}(n^{-1/3})$ (Assumption C2 below). Assume that $%
\{(X_{i},W_{i},S_{i},D_{i})\}_{i=1}^{n}$ is a random sample from the joint
distribution of $(X,W,S,D)$. Let\ $\hat{U}_{k}\equiv \frac{1}{n}%
\sum_{i=1}^{n}1\{X_{i}^{\top }\hat{\theta}\leq X_{k}^{\top }\hat{\theta}\}$
and define%
\begin{equation*}
\hat{\mu}(\hat{U}_{k})\equiv \frac{\sum_{i=1}^{n}D_{i}W_{i}K_{h}(\hat{U}_{i}-%
\hat{U}_{k})}{\sum_{i=1}^{n}D_{i}K_{h}(\hat{U}_{i}-\hat{U}_{k})},
\end{equation*}%
as an estimator of $\mu _{0}(U_{0})$, where $K_{h}(u)\equiv K(u/h)/h,$ $K:%
\mathbf{R}\rightarrow \mathbf{R}$ is a kernel function, and $h$ is a
bandwidth parameter. The estimator is a symmetrized nearest neighborhood
(SNN) estimator. Symmetrized nearest neighborhood estimation is a variant of
nearest neighborhood estimation originated by Fix and Hodges (1951), and
analyzed and expanded by Stone (1977). Robinson (1987) introduced $k$%
-nearest neighborhood estimation in semiparametric models in the estimation
of conditional heteroskedasticity of unknown form. The symmetrized nearest
neighborhood estimation that this paper uses was proposed by Yang (1981) and
further studied by Stute (1984).

Define%
\begin{equation}
\hat{a}(\hat{\theta})\equiv \frac{1}{\sum_{i=1}^{n}D_{i}}\sum_{i=1}^{n}D_{i}%
\{S_{i}\cdot \varphi (\hat{\mu}(\hat{U}_{i}))\}\text{.}  \label{ne}
\end{equation}%
The estimator $\hat{a}(\hat{\theta})$ is a sample analogue of $a(\theta
_{0}) $, where the conditional expectations are replaced by the
nonparametric estimators and unconditional expectations by the sample mean.

Suppose that we are given a consistent estimator $\hat{b}$ of $b_{0}$
(Assumption G1(iii) below). Then, our estimator takes the following form:%
\begin{equation}
\hat{\beta}=H(\hat{a}(\hat{\theta}),\hat{b}).  \label{b}
\end{equation}%
We make the following assumptions. The assumptions are divided into two
groups. The first group of assumptions (denoted by Assumptions C1-C3) are
commonly assumed throughout the examples when we discuss them later. On the
other hand, the second group of assumptions (denoted by Assumptions G1-G2)
are the ones for which sufficient conditions will be provided later when we
discuss the examples.\bigskip

\noindent \textsc{Assumption C1} \textsc{:}\textbf{\ }(i) For some $%
\varepsilon >0$, $P\{D=1|U_{0}=u\}>\varepsilon $ for all $u\in \lbrack 0,1]$.

\noindent (ii) There exists $\varepsilon >0$ such that for each $\theta \in
B(\theta _{0};\varepsilon ),$ (a) $X^{\top }\theta $ is continuous and its
conditional density function given $D=1$ is bounded uniformly over $\theta
\in B(\theta _{0};\varepsilon )$ and bounded away from zero on the interior
of its support uniformly over $\theta \in B(\theta _{0};\varepsilon )$, and
(b) the set $\{x^{\prime }\theta :\theta \in B(\theta _{0};\varepsilon
),x\in S_{m}\}$ is an interval of finite length for all $1\leq m\leq M$%
.\bigskip

\noindent \textsc{Assumption C2} \textsc{:} $||\hat{\theta}-\theta
_{0}||=O_{P}(n^{-1/3}).$\bigskip

\noindent \textsc{Assumption C3} \textsc{:}\textbf{\ }(i) $K(\cdot )$ is
bounded, nonnegative, symmetric, compact supported, twice continuously
differentiable with bounded derivatives on the interior of the support,\ and 
$\int K(t)dt=1$.

\noindent (ii) $n^{1/2}h^{3}+n^{-1/2}h^{-2}(-\log h)\rightarrow 0.$\bigskip

Assumption C1(ii)(a) excludes the case where $\theta _{0}=0$. Assumption
C1(ii)(b) is satisfied when $S_{m}$ is bounded and convex. We can weaken
Assumption C1(ii) by replacing it with certain tail conditions of $X$ or $%
X^{\top }\theta $ at the expense of a more complicated exposition.
Assumption C2 allows $\hat{\theta}$ to be either $\sqrt{n}$-consistent or $%
\sqrt[3]{n}$-consistent. Assumption C3 concerns the kernel and the
bandwidth. Assumption C3(i) is satisfied, for example, by a quartic kernel:\ 
$K(u)=(15/16)(1-u^{2})^{2}1\{|u|\leq 1\}.$\bigskip

\noindent \textsc{Assumption G1} \textsc{:} (i) For $p\geq 4,\ $sup$_{x\in 
\mathcal{S}_{X}}\mathbf{E}\left[ ||W||^{p}|X=x\right] +$sup$_{x\in \mathcal{S%
}_{X}}\mathbf{E}\left[ ||S||^{p}|X=x\right] <\infty .$

\noindent (ii) (a) $\varphi $ is twice continuously differentiable on the
interior of the support of $\mathbf{E}[W|X]$ with derivatives bounded on the
support of $\mathbf{E}[W|X]$, and (b) there exists $\eta >0$ such that for
all $b\in B(b_{0};\eta )$, $H(\cdot ,b)$ is continuously differentiable at $%
a=a(\theta _{0})$ and the derivative $\partial H(a,b)/\partial a\ $is
continuous at $(a(\theta _{0}),b_{0})$.

\noindent (iii) $\hat{b}=b_{0}+o_{P}(1).$\bigskip

\noindent \textsc{Assumption G2} \textsc{:}$\ $(i) For each $m=1,\cdot \cdot
\cdot ,M,$ both $\mathbf{E}[S|X_{1}=\cdot ,(X_{2},D)=(x_{m},1)]$ and $%
\mathbf{E}[W|X_{1}=\cdot ,(X_{2},D)=(x_{m},1)]$ are Lipschitz continuous.

\noindent (ii) $\mathbf{E}[W|U_{\theta }=\cdot ]$ is twice continuously
differentiable with derivatives bounded uniformly over $\theta \in B(\theta
_{0};\varepsilon )$ with some $\varepsilon >0.$\bigskip

Assumption G1(i) requires moment conditions with $p\geq 4$. Assumption
G1(ii) is easy to check, because $\varphi $ is explicitly known in many
examples. Assumption G1(iii) requires that $\hat{b}$ be a consistent
estimator of $b_{0}$. As we will see from the examples, a $\sqrt{n}$%
-consistent estimator for $b_{0}$ is typically available. The smoothness
conditions in Assumptions G2(i) and (ii) are often used in the literature of
nonparametric estimation.\bigskip

\noindent \textsc{Theorem 1:} \textit{Suppose that Assumptions C1--C3 and
G1-G2 hold and that }$\sqrt{n}(\tilde{\beta}-\beta )\overset{d}{\rightarrow }%
N(0,V)$\textit{\ for some positive definite matrix }$V$\textit{. Then}%
\begin{equation*}
\sqrt{n}(\hat{\beta}-\beta )\overset{d}{\rightarrow }N(0,V).
\end{equation*}

\noindent \textsc{Remarks 1} \textsc{:} The asymptotic covariance matrix $V$
in Theorem 1 is the same asymptotic covariance matrix that one would have
obtained had $\theta _{0}$ been used instead of $\hat{\theta}$. Therefore,
the estimation error in $\hat{\theta}$ does not affect the asymptotic
distribution of $\hat{\beta}$. When the nuisance parameter estimator $\hat{%
\theta}$ is $\sqrt{n}$-consistent, such a phenomenon has been observed to
arise in other contexts (e.g. Song (2009)). To the best of the author's
knowledge, there has not been a literature that shows a similar phenomenon
even when $\hat{\theta}$ is $n^{1/3}$-consistent.\bigskip

\noindent \textsc{2:} The computation of $V$ such that $\sqrt{n}(\tilde{\beta%
}-\beta )\overset{d}{\rightarrow }N(0,V)$ can be done using the standard
procedure. (e.g. Newey and McFadden (1994)). Section 3.2 below derives the
asymptotic covariance matrix $V$ for the examples in Section 2.2. For the
derivation, one does not need to rely on the form (\ref{param}). Writing $%
\beta _{0}$ into the form (\ref{param}) is done only to ensure that Theorem
1 is applicable.\bigskip

\noindent \textsc{3:} The proof of Theorem 1 uses a Bahadur representation
of sample linear functionals of SNN\ estimators that is established in the
Appendix. In fact, the representation can also be used to derive the
asymptotic covariance matrix $V$, and is useful in various specification
tests or estimation for semiparametric models.\bigskip 

\noindent \textsc{4:} Theorem 1 implies that there exists a simple bootstrap
procedure for $\hat{\beta}$ that is asymptotically valid, even if the
first-step estimator $\hat{\theta}$ follows cube-root asymptotics. This is
interesting given that nonparametric bootstrap fails for $\hat{\theta}$.
(Abrevaya and Huang (2005)). Since there is no clear advantage of using this
bootstrap over the asymptotic covariance matrix of Theorem 1, this paper
omits the details.

\subsection{Examples Revisited}

In this section, we revisit the examples discussed in Section 2.2. In each
example, we first provide sufficient conditions that yield Assumptions
G1-G2. (Recall that Assumptions C1-C3 are made commonly in these examples.)
Then we show how we construct an estimator of $\beta _{0}$ in detail.
Finally, we present the asymptotic distribution of $\hat{\beta}$ along with
the explicit asymptotic covariance matrix formula.

\subsubsection{Example 1: Sample Selection Models with Conditional Median
Restrictions}

In this example, Assumptions G1-G2 are translated into the following
conditions.\bigskip

\noindent \textsc{Assumption SS1} \textsc{:} For $p\geq 4,\ $sup$_{x\in 
\mathcal{S}_{X}}\mathbf{E}\left[ |Y|^{p}|X=x\right] +$ sup$_{x\in \mathcal{S}%
_{X}}\mathbf{E}\left[ ||Z||^{p}|X=x\right] <\infty $.\bigskip

\noindent \textsc{Assumption SS2} \textsc{:} (i) For each $m=1,\cdot \cdot
\cdot ,M,$ both $\mathbf{E}[Y|X_{1}=\cdot ,(X_{2},D)=(x_{m},1)]$ and $%
\mathbf{E}[Z|X_{1}=\cdot ,(X_{2},D)=(x_{m},1)]$ are Lipschitz continuous.

\noindent (ii) $\mathbf{E}[Y|U_{\theta }=\cdot ]$ and $\mathbf{E}%
[Z|U_{\theta }=\cdot ]$ are twice continuously differentiable with
derivatives bounded uniformly over $\theta \in B(\theta _{0};\varepsilon )$
with some $\varepsilon >0.$\bigskip

Since $\varphi $ that constitutes $a(\theta )$ is an identity map in this
example, Assumption G1(ii)(a) is already fulfilled by Assumptions SS1 and
SS2(ii).

Let us consider an estimator of $\beta _{0}$ in the sample selection models
in Example 1. With $\hat{U}_{k}$ as defined previously, let%
\begin{equation}
\hat{\mu}_{Y}(\hat{U}_{k})\equiv \frac{\sum_{i=1}^{n}D_{i}Y_{i}K_{h}(\hat{U}%
_{i}-\hat{U}_{k})}{\sum_{i=1}^{n}D_{i}K_{h}(\hat{U}_{i}-\hat{U}_{k})}\text{
and }\hat{\mu}_{Z}(\hat{U}_{k})\equiv \frac{\sum_{i=1}^{n}D_{i}Z_{i}K_{h}(%
\hat{U}_{i}-\hat{U}_{k})}{\sum_{i=1}^{n}D_{i}K_{h}(\hat{U}_{i}-\hat{U}_{k})}%
\text{.}  \label{muYZ}
\end{equation}%
Using $\hat{\mu}_{Y}(\hat{U}_{k})$ and $\hat{\mu}_{Z}(\hat{U}_{k})$, we
define%
\begin{eqnarray*}
\hat{S}_{ZZ} &\equiv &\frac{1}{\sum_{i=1}^{n}D_{i}}\sum_{i=1}^{n}(Z_{i}-\hat{%
\mu}_{Z}(\hat{U}_{i}))(Z_{i}-\hat{\mu}_{Z}(\hat{U}_{i}))^{\top }D_{i}\text{
and} \\
\hat{S}_{ZY} &\equiv &\frac{1}{\sum_{i=1}^{n}D_{i}}\sum_{i=1}^{n}(Z_{i}-\hat{%
\mu}_{Z}(\hat{U}_{i}))(Y_{i}-\hat{\mu}_{Y}(\hat{U}_{i}))D_{i},
\end{eqnarray*}%
which are estimated versions of $S_{ZZ}(\theta _{0})$ and $S_{ZY}(\theta
_{0})$. An estimator of $\beta _{0}$ is given by%
\begin{equation}
\hat{\beta}\equiv \hat{S}_{ZZ}^{-1}\cdot \hat{S}_{ZY}.  \label{beta}
\end{equation}%
This is the estimator proposed by Robinson (1988), except that we have a
single-index in the conditional expectations. We also let $\tilde{\beta}$ be 
$\hat{\beta}$ except that $\hat{\theta}$ is replaced by $\theta _{0}$.

Suppose that Assumptions C1--C3 and SS1-SS2 hold and that $\sqrt{n}(\tilde{%
\beta}-\beta _{0})\overset{d}{\rightarrow }N(0,V_{SS})$ for some positive
definite matrix $V_{SS}$. Then by Theorem 1,%
\begin{equation*}
\sqrt{n}(\hat{\beta}-\beta _{0})\overset{d}{\rightarrow }N(0,V_{SS}).
\end{equation*}%
The computation of $V_{SS}$ can be done in a standard manner. Under
regularity conditions, the asymptotic variance $V_{SS}$ takes the following
form:\textit{\ }$V_{SS}=S_{ZZ}^{-1}(\theta _{0})\Omega S_{ZZ}^{-1}(\theta
_{0})$ with $\sigma ^{2}(U_{0})\equiv Var(v|U_{0},D=1),$%
\begin{equation*}
\Omega \equiv \mathbf{E}\left[ \sigma ^{2}(U_{0})(Z-\mathbf{E}%
[Z|D=1,U_{0}])(Z-\mathbf{E}[Z|D=1,U_{0}])^{\top }|D=1\right] /P_{1}.
\end{equation*}%
The derivation can be obtained by using the Bahadur representation in the
appendix (Lemma B3).

\subsubsection{Example 2: Single-Index Matching Estimators of Treatment
Effects on the Treated}

We introduce translations of Assumptions G1-G2 in this example.\bigskip

\noindent \textsc{Assumption SM1} \textsc{:} For $p\geq 4,\ $sup$_{x\in 
\mathcal{S}_{X}}\mathbf{E}\left[ |Y|^{p}|X=x\right] <\infty .$\bigskip

\noindent \textsc{Assumption SM2} \textsc{:}(i) For each $m=1,\cdot \cdot
\cdot ,M,$ both $P\{Z=1|X_{1}=\cdot ,X_{2}=x_{m}\}$ and $\mathbf{E}%
[Y|X_{1}=\cdot ,X_{2}=x_{m}]$ are Lipschitz continuous.

\noindent (ii) $P\{Z=1|U_{\theta }=\cdot \}$ and $\mathbf{E}[Y|U_{\theta
}=\cdot ]$ are twice continuously differentiable with derivatives bounded
uniformly over $\theta \in B(\theta _{0};\varepsilon )$ with some $%
\varepsilon >0.$\bigskip

To construct an estimator of the average treatment effect on the treated
based on the single-index matching, we first define%
\begin{eqnarray*}
\hat{\mu}_{(1-Z)Y}(\hat{U}_{k}) &\equiv &\frac{%
\sum_{i=1}^{n}(1-Z_{i})Y_{i}K_{h}(\hat{U}_{i}-\hat{U}_{k})}{%
\sum_{i=1}^{n}K_{h}(\hat{U}_{i}-\hat{U}_{k})}\text{ and} \\
\hat{P}(\hat{U}_{k}) &\equiv &\frac{\sum_{i=1}^{n}Z_{i}K_{h}(\hat{U}_{i}-%
\hat{U}_{k})}{\sum_{i=1}^{n}K_{h}(\hat{U}_{i}-\hat{U}_{k})}.
\end{eqnarray*}%
Then, the sample analogue principle suggests%
\begin{equation*}
\hat{\beta}=\frac{1}{\sum_{i=1}^{n}Z_{i}}\sum_{k=1}^{n}Z_{k}\left\{ Y_{k}-%
\frac{\hat{\mu}_{(1-Z)Y}(\hat{U}_{k})}{1-\hat{P}(\hat{U}_{k})}\right\} .
\end{equation*}%
If we define%
\begin{equation*}
\hat{\mu}(\hat{U}_{k})=\frac{\sum_{i=1}^{n}(1-Z_{i})Y_{i}K_{h}(\hat{U}_{i}-%
\hat{U}_{k})}{\sum_{i=1}^{n}(1-Z_{i})K_{h}(\hat{U}_{i}-\hat{U}_{k})},
\end{equation*}%
we can rewrite the estimator as%
\begin{equation*}
\hat{\beta}=\frac{1}{\sum_{i=1}^{n}Z_{i}}\sum_{k=1}^{n}Z_{k}\{Y_{k}-\hat{\mu}%
(\hat{U}_{k})\}.
\end{equation*}%
This takes precisely the same form as the propensity score matching
estimators of Heckman, Ichimura, and Todd (1998), except that instead of
propensity score matching, the estimator uses single-index matching.

As before, we let $\tilde{\beta}$ be $\hat{\beta}$ except that $\hat{\theta}$
is replaced by $\theta _{0}$. Suppose that Assumptions C1--C3 and SM1-SM2
hold and that $\sqrt{n}(\tilde{\beta}-\beta _{0})\overset{d}{\rightarrow }%
N(0,V_{SM})$ for some positive definite matrix $V_{SM}$. Then, by Theorem 1,%
\begin{equation*}
\sqrt{n}(\hat{\beta}-\beta _{0})\overset{d}{\rightarrow }N(0,V_{SM}).
\end{equation*}%
Under regularity conditions, the asymptotic variance $V_{SM}$ takes the
following form: with $\mu _{d}(U_{0})=\mathbf{E}[Y|U_{0},Z=d]$ and $%
P_{d}=P\{Z=d\}$ for $d\in \{0,1\},$%
\begin{eqnarray*}
V_{SM} &=&\mathbf{E}\left[ \left( Y-\mu _{1}(U_{0})\right) ^{2}|Z=1\right]
/P_{1} \\
&&+\mathbf{E}\left[ \left( Y-\mu _{0}(U_{0})\right)
^{2}P^{2}(U_{0})/(1-P(U_{0}))^{2}|Z=0\right] (1-P_{1})/P_{1}^{2} \\
&&+Var\left( \mu _{1}(U_{0})-\mu _{0}(U_{0})|Z=1\right) /P_{1}.
\end{eqnarray*}

\section{A Monte Carlo Simulation Study}

In this section, we present and discuss some Monte Carlo simulation results.
We consider the following data generating process. Let%
\begin{equation*}
Z_{i}=U_{1i}-(\eta _{1i}/2)\mathbf{1}\text{ and\ }X_{i}=U_{2i}-\eta _{i}/2,
\end{equation*}%
where $U_{1i}$ is an i.i.d. random vector in $\mathbf{R}^{3}\ $constituted
by independent random variables with uniform distribution on $[0,1]$, $%
\mathbf{1}$ is a $3$ dimensional vector of ones, $U_{2i}$ and $\eta _{i}$
are random vectors in $\mathbf{R}^{k}\ $with entries equal to i.i.d. random
variables of uniform distribution on $[0,1].$ The dimension $k$ is chosen
from $\{3,6\}.\ $The random variable $\eta _{1i}$ is the first component of $%
\eta _{i}.\ $Then, the selection mechanism is defined as%
\begin{equation*}
D_{i}=1\{X_{i}^{\top }\theta _{0}+\varepsilon _{i}\geq 0\},
\end{equation*}%
where $\varepsilon _{i}$ follows the distribution of $T_{i}\cdot \varphi
(X_{i}^{\top }\theta _{0})+e_{i}$, with $e_{i}\sim N(0,1)$, and $T_{i}$ and $%
\varphi (\cdot )$ are chosen as follows:%
\begin{eqnarray*}
\text{Specification AN} &\text{: }&T_{i}\sim N(0,1),\ \varphi =2\Phi
(z^{2}+|z|) \\
\text{Specification AT} &\text{: }&T_{i}\sim t_{1},\ \varphi =2\Phi
(z^{2}+|z|), \\
\text{Specification BN} &\text{:}&T_{i}\sim N(0,1)+e_{i},\ \varphi =\exp
(z-1) \\
\text{Specification BT} &\text{: }&T_{i}\sim t_{1}+e_{i},\ \varphi =\exp
(z-1),
\end{eqnarray*}%
where $t_{1}$ denotes the $t$-distribution with degree of freedom 1 and $%
\Phi $ denotes the standard normal CDF.

Hence the selection mechanism has errors that are conditionally
heteroskedastic, and in the case of DGPs with AT and BT, heavy tailed. We
define the latent outcome $Y_{i}^{\ast }$ as follows:%
\begin{equation*}
Y_{i}^{\ast }=Z_{i}^{\top }\beta _{0}+v_{i},
\end{equation*}%
where $v_{i}\sim (2\zeta _{i}+e_{i})\times 2\Phi \left( (X_{i}^{\top }\theta
_{0})^{2}+|X_{i}^{\top }\theta _{0}|\right) $ and $\zeta _{i}\sim N(0,1)$
independent of the other random variables. Therefore, $v_{i}$ in the outcome
equation and $\varepsilon _{i}$ in the selection equation are correlated, so
that the data generating process admits the sample selection bias. We set $%
\theta _{0}$ to be the vector of $2$'s and $\beta _{0}=[2,2,2]^{\top }.$ In
the simulation studies we estimated $\theta _{0}$ by using the maximum score
estimation to obtain $\hat{\theta}$.

We compare the performances of the two estimators of $\beta _{0}$, $\hat{%
\beta}(\hat{\theta})$ ("Plug-in $\hat{\theta}$") and $\hat{\beta}(\theta
_{0})$ ("Plug-in $\theta _{0}$") in terms of mean absolute deviation (MAE)
and mean squared error (MSE). Bandwidths for the estimation of $\mathbf{E}%
[Y_{i}|X_{i}^{\top }\theta _{0},D_{i}=1]$ and $\mathbf{E}[Z_{i}|X_{i}^{\top
}\theta _{0},D_{i}=1]$ were chosen separately using a least-squares
cross-validation method. If the role of the sample selection bias were
already marginal, the estimation error effect of $\hat{\theta}$ would be
small accordingly, preventing us from discerning the negligibility of the
estimation error effect of $\hat{\theta}$ from the negligible sample
selection bias. Hence, we also report the results from the estimation of $%
\beta $ that ignores the sample selection bias (w/o BC: Without (Sample
Selection) Bias Correction).

Table 1 reports the average of MAEs and MSEs of estimators for the
individual components of $\beta _{0}$. It shows that the performance of the
estimators remains similar regardless of whether $\theta _{0}$ is used or $%
\hat{\theta}$ is used. When the sample size is increased from 300 to 500,
the estimators perform better as expected. The negligibility of the effect
of the estimation error in $\hat{\theta}$ is not due to inherently weak
sample selection bias, as it is evident when we compare the results with
those from the estimators that ignore the sample selection bias (w/o
BC).\pagebreak \bigskip

\begin{center}
{\small Table 1: The Performance of the Estimators in Terms of MAE and RMSE
(Specification A)\bigskip }

\begin{tabular}{lll|lll|lll}
\hline\hline
&  &  &  & ${\small k=3}$ &  &  & ${\small k=6}$ &  \\ 
& {\small Specification} &  & {\small Plug-in }$\theta _{0}$ & {\small %
Plug-in }$\hat{\theta}$ & {\small w/o BC} & {\small Plug-in }$\theta _{0}$ & 
{\small Plug-in }$\hat{\theta}$ & {\small w/o BC} \\ \hline\hline
& {\small Spec. AN} & {\small MAE} & {\small 0.6911} & {\small 0.6909} & 
{\small 0.7837} & {\small 0.6458} & {\small 0.6453} & {\small 0.6436} \\ 
${\small n=300}$ &  & {\small RMSE} & {\small 2.2533} & {\small 2.2539} & 
{\small 2.8485} & {\small 1.9634} & {\small 1.9621} & {\small 1.9479} \\ 
\cline{2-9}
& {\small Spec. AT} & {\small MAE} & {\small 0.7345} & {\small 0.7351} & 
{\small 0.7729} & {\small 0.6913} & {\small 0.6918} & {\small 0.6739} \\ 
&  & {\small RMSE} & {\small 2.5437} & {\small 2.5457} & {\small 2.7955} & 
{\small 2.2408} & {\small 2.2446} & {\small 2.1316} \\ \hline
& {\small Spec. AN} & {\small MAE} & {\small 0.5327} & {\small 0.5328} & 
{\small 0.6717} & {\small 0.4965} & {\small 0.4966} & {\small 0.5122} \\ 
${\small n=500}$ &  & {\small RMSE} & {\small 1.3428} & {\small 1.3432} & 
{\small 2.0406} & {\small 1.1615} & {\small 1.1620} & {\small 1.2270} \\ 
\cline{2-9}
& {\small Spec. AT} & {\small MAE} & {\small 0.5658} & {\small 0.5654} & 
{\small 0.6360} & {\small 0.5308} & {\small 0.5310} & {\small 0.5316} \\ 
&  & {\small RMSE} & {\small 1.5154} & {\small 1.5134} & {\small 1.8833} & 
{\small 1.3256} & {\small 1.3263} & {\small 1.3318} \\ \hline
& {\small Spec. AN} & {\small MAE} & {\small 0.3766} & {\small 0.3765} & 
{\small 0.5765} & {\small 0.3475} & {\small 0.3475} & {\small 0.3880} \\ 
${\small n=1000}$ &  & {\small RMSE} & {\small 0.6693} & {\small 0.6696} & 
{\small 1.4206} & {\small 0.5707} & {\small 0.5707} & {\small 0.6993} \\ 
\cline{2-9}
& {\small Spec. AT} & {\small MAE} & {\small 0.3981} & {\small 0.3980} & 
{\small 0.5182} & {\small 0.3734} & {\small 0.3734} & {\small 0.3982} \\ 
&  & {\small RMSE} & {\small 0.7455} & {\small 0.7449} & {\small 1.2094} & 
{\small 0.6570} & {\small 0.6572} & {\small 0.7368} \\ \hline
\end{tabular}

\bigskip \bigskip 

{\small Table 2: The Performance of the Estimators in Terms of MAE and RMSE
(Specification B)\bigskip }

\begin{tabular}{lll|lll|lll}
\hline\hline
&  &  &  & ${\small k=3}$ &  &  & ${\small k=6}$ &  \\ 
& {\small Specification} &  & {\small Plug-in }$\theta _{0}$ & {\small %
Plug-in }$\hat{\theta}$ & {\small w/o BC} & {\small Plug-in }$\theta _{0}$ & 
{\small Plug-in }$\hat{\theta}$ & {\small w/o BC} \\ \hline\hline
& {\small Spec. BN} & {\small MAE} & {\small 0.6707} & {\small 0.6710} & 
{\small 1.4572} & {\small 0.7017} & {\small 0.7020} & {\small 1.4565} \\ 
${\small n=300}$ &  & {\small RMSE} & {\small 2.1182} & {\small 2.1211} & 
{\small 8.0981} & {\small 2.3237} & {\small 2.3277} & {\small 8.1755} \\ 
\cline{2-9}
& {\small Spec. BT} & {\small MAE} & {\small 0.7209} & {\small 0.7212} & 
{\small 1.2461} & {\small 0.7563} & {\small 0.7561} & {\small 1.2164} \\ 
&  & {\small RMSE} & {\small 2.4474} & {\small 2.4489} & {\small 6.3876} & 
{\small 2.7029} & {\small 2.7050} & {\small 6.2357} \\ \hline
& {\small Spec. BN} & {\small MAE} & {\small 0.5260} & {\small 0.5260} & 
{\small 1.4355} & {\small 0.5504} & {\small 0.5515} & {\small 1.4331} \\ 
${\small n=500}$ &  & {\small RMSE} & {\small 1.2997} & {\small 1.2990} & 
{\small 7.2972} & {\small 1.4298} & {\small 1.4350} & {\small 7.3290} \\ 
\cline{2-9}
& {\small Spec. BT} & {\small MAE} & {\small 0.5649} & {\small 0.5649} & 
{\small 1.1998} & {\small 0.5867} & {\small 0.5870} & {\small 1.1574} \\ 
&  & {\small RMSE} & {\small 1.5078} & {\small 1.5069} & {\small 5.5043} & 
{\small 1.6318} & {\small 1.6338} & {\small 5.2330} \\ \hline
& {\small Spec. BN} & {\small MAE} & {\small 0.3870} & {\small 0.3869} & 
{\small 1.4258} & {\small 0.4174} & {\small 0.4184} & {\small 1.4317} \\ 
${\small n=1000}$ &  & {\small RMSE} & {\small 0.7040} & {\small 0.7040} & 
{\small 6.6726} & {\small 0.8148} & {\small 0.8190} & {\small 6.7706} \\ 
\cline{2-9}
& {\small Spec. BT} & {\small MAE} & {\small 0.4053} & {\small 0.4052} & 
{\small 1.1746} & {\small 0.4230} & {\small 0.4235} & {\small 1.1291} \\ 
&  & {\small RMSE} & {\small 0.7721} & {\small 0.7709} & {\small 4.7835} & 
{\small 0.8444} & {\small 0.8469} & {\small 4.5057} \\ \hline
\end{tabular}

\bigskip 
\end{center}

In Tables 3 and 4, the finite sample coverage probabilities of the
confidence sets based on the asymptotic normal distribution are reported.
These tables report only the coverage probabilities of the first component
of the estimators of $\beta _{0}$. The performance of the remaining
components was similar. In Table 3, the results were obtained from
Specification A, and in Table 4, from Specification B. Recall that
Specification B is associated with a more severe selection bias than
Specification A as we saw from Tables 1-2.\pagebreak \bigskip

\begin{center}
{\small Table 3: The Performance of the Confidence Intervals (Specification
A)\bigskip }

\begin{tabular}{ccc|cc|cc}
\hline\hline
&  &  & ${\small k=3}$ &  & ${\small k=6}$ &  \\ 
& {\small Specification} & {\small Nom. Cov. Prob.} & {\small Plug-in }$%
\theta _{0}$ & {\small Plug-in }$\hat{\theta}$ & {\small Plug-in }$\theta
_{0}$ & {\small Plug-in }$\hat{\theta}$ \\ \hline\hline
&  & {\small 99\%} & {\small 0.9877} & {\small 0.9881} & {\small 0.9875} & 
{\small 0.9877} \\ 
& {\small Spec. AN} & {\small 95\%} & {\small 0.9449} & {\small 0.9444} & 
{\small 0.9487} & {\small 0.9468} \\ 
${\small n=300}$ &  & {\small 90\%} & {\small 0.8938} & {\small 0.8930} & 
{\small 0.8972} & {\small 0.8947} \\ \cline{2-7}
&  & {\small 99\%} & {\small 0.9858} & {\small 0.9866} & {\small 0.9871} & 
{\small 0.9861} \\ 
& {\small Spec.AT} & {\small 95\%} & {\small 0.9416} & {\small 0.9398} & 
{\small 0.9429} & {\small 0.9427} \\ 
&  & {\small 90\%} & {\small 0.8866} & {\small 0.8878} & {\small 0.8900} & 
{\small 0.8900} \\ \hline
&  & {\small 99\%} & {\small 0.9886} & {\small 0.9891} & {\small 0.9879} & 
{\small 0.9880} \\ 
& {\small Spec. AN} & {\small 95\%} & {\small 0.9480} & {\small 0.9484} & 
{\small 0.9456} & {\small 0.9453} \\ 
${\small n=500}$ &  & {\small 90\%} & {\small 0.8988} & {\small 0.8980} & 
{\small 0.8957} & {\small 0.8978} \\ \cline{2-7}
&  & {\small 99\%} & {\small 0.9859} & {\small 0.9864} & {\small 0.9878} & 
{\small 0.9883} \\ 
& {\small Spec. AT} & {\small 95\%} & {\small 0.9445} & {\small 0.9449} & 
{\small 0.9466} & {\small 0.9472} \\ 
&  & {\small 90\%} & {\small 0.8940} & {\small 0.8950} & {\small 0.8935} & 
{\small 0.8958} \\ \hline
&  & {\small 99\%} & {\small 0.9891} & {\small 0.9884} & {\small 0.9909} & 
{\small 0.9910} \\ 
& {\small Spec. AN} & {\small 95\%} & {\small 0.9463} & {\small 0.9468} & 
{\small 0.9513} & {\small 0.9508} \\ 
${\small n=1000}$ &  & {\small 90\%} & {\small 0.8956} & {\small 0.8964} & 
{\small 0.9030} & {\small 0.9031} \\ \cline{2-7}
&  & {\small 99\%} & {\small 0.9858} & {\small 0.9862} & {\small 0.9898} & 
{\small 0.9899} \\ 
& {\small Spec. AT} & {\small 95\%} & {\small 0.9435} & {\small 0.9447} & 
{\small 0.9488} & {\small 0.9474} \\ 
&  & {\small 90\%} & {\small 0.8953} & {\small 0.8969} & {\small 0.8993} & 
{\small 0.9005} \\ \hline
\end{tabular}

\bigskip \bigskip 

{\small Table 4: The Performance of the Confidence Intervals (Specification
B)\bigskip }

\begin{tabular}{ccc|cc|cc}
\hline\hline
&  &  & ${\small k=3}$ &  & ${\small k=6}$ &  \\ 
& {\small Specification} & {\small Nom. Cov. Prob.} & {\small Plug-in }$%
\theta _{0}$ & {\small Plug-in }$\hat{\theta}$ & {\small Plug-in }$\theta
_{0}$ & {\small Plug-in }$\hat{\theta}$ \\ \hline\hline
&  & {\small 99\%} & {\small 0.9832} & {\small 0.9829} & {\small 0.9824} & 
{\small 0.9823} \\ 
& {\small Spec. BN} & {\small 95\%} & {\small 0.9334} & {\small 0.9340} & 
{\small 0.9339} & {\small 0.9345} \\ 
${\small n=300}$ &  & {\small 90\%} & {\small 0.8852} & {\small 0.8837} & 
{\small 0.8777} & {\small 0.8796} \\ \cline{2-7}
&  & {\small 99\%} & {\small 0.9851} & {\small 0.9853} & {\small 0.9833} & 
{\small 0.9827} \\ 
& {\small Spec.BT} & {\small 95\%} & {\small 0.9382} & {\small 0.9388} & 
{\small 0.9377} & {\small 0.9384} \\ 
&  & {\small 90\%} & {\small 0.8898} & {\small 0.8903} & {\small 0.8844} & 
{\small 0.8851} \\ \hline
&  & {\small 99\%} & {\small 0.9834} & {\small 0.9833} & {\small 0.9809} & 
{\small 0.9805} \\ 
& {\small Spec. BN} & {\small 95\%} & {\small 0.9353} & {\small 0.9357} & 
{\small 0.9295} & {\small 0.9289} \\ 
${\small n=500}$ &  & {\small 90\%} & {\small 0.8825} & {\small 0.8841} & 
{\small 0.8770} & {\small 0.8744} \\ \cline{2-7}
&  & {\small 99\%} & {\small 0.9844} & {\small 0.9844} & {\small 0.9828} & 
{\small 0.9826} \\ 
& {\small Spec. BT} & {\small 95\%} & {\small 0.9399} & {\small 0.9402} & 
{\small 0.9360} & {\small 0.9363} \\ 
&  & {\small 90\%} & {\small 0.8829} & {\small 0.8837} & {\small 0.8838} & 
{\small 0.8815} \\ \hline
&  & {\small 99\%} & {\small 0.9797} & {\small 0.9794} & {\small 0.9765} & 
{\small 0.9770} \\ 
& {\small Spec. BN} & {\small 95\%} & {\small 0.9264} & {\small 0.9276} & 
{\small 0.9156} & {\small 0.9143} \\ 
${\small n=1000}$ &  & {\small 90\%} & {\small 0.8679} & {\small 0.8680} & 
{\small 0.8511} & {\small 0.8505} \\ \cline{2-7}
&  & {\small 99\%} & {\small 0.9859} & {\small 0.9858} & {\small 0.9833} & 
{\small 0.9830} \\ 
& {\small Spec. BT} & {\small 95\%} & {\small 0.9364} & {\small 0.9367} & 
{\small 0.9323} & {\small 0.9310} \\ 
&  & {\small 90\%} & {\small 0.8811} & {\small 0.8820} & {\small 0.8760} & 
{\small 0.8750} \\ \hline
\end{tabular}

\bigskip 
\end{center}

First, observe that the performances between the estimator using true
parameter $\theta _{0}$ and the estimator using its estimator $\hat{\theta}$
is almost negligible in finite samples, as expected from the asymptotic
theory. This is true regardless of whether we use three or six covariates.
In Specification B, the confidence sets tend to undercover the true
parameter, as compared to Specification A. Nevertheless, the difference in
coverage probabilities between the estimator using $\theta _{0}$ and the
estimator using $\hat{\theta}$ is still very negligible. Finally, the
performances do not show much difference with regard to the heavy tailedness
of the error distribution in the selection equation, as seen from comparing
results between Specifications AN and AT or between Specifications BN and BT.

\section{Empirical Application: Female Labor Supply from NLSY79}

In this section, we illustrate the proposal of this paper by estimating a
simple female labor supply model:%
\begin{eqnarray*}
h_{i} &=&\beta _{0}+\log (w_{i})\beta _{1}+Z_{2i}^{\top }\beta
_{4}+\varepsilon _{i}\text{ and} \\
D_{i} &=&1\left\{ X_{i}^{\top }\theta _{0}\geq \eta _{i}\right\} ,
\end{eqnarray*}%
where $h_{i}$ denotes hours that the $i$-th female worker worked, $w_{i}$
her hourly wage, $Z_{2i}$ denotes other demographic variables. The following
table shows different specifications that this study used.\bigskip

\begin{center}
\bigskip 

Table 5 : Variables used for $Z_{2i}$ and $X_{i}$\bigskip

$%
\begin{tabular}{ll|l}
\hline\hline
&  & \ \ \ \ \ \ Variables Used \\ \hline\hline
Specification I & $Z_{2i}$ & 
\begin{tabular}{l}
nonwife income, age, schooling \\ 
\# kids w/ age 0-5,\ \# kids w/ age 6-18,%
\end{tabular}
\\ \cline{2-3}
& $X_{i}$ & 
\begin{tabular}{l}
mother and father's schooling \\ 
age, schooling%
\end{tabular}
\\ \hline
Specification II & $Z_{2i}$ & 
\begin{tabular}{l}
nonwife income, age, schooling \\ 
\# kids w/ age 0-5,\ \# kids w/ age 6-18,%
\end{tabular}
\\ \cline{2-3}
& $X_{i}$ & 
\begin{tabular}{l}
mother and father's schooling \\ 
age, schooling, and household income%
\end{tabular}
\\ \hline
Specification III & $Z_{2i}$ & 
\begin{tabular}{l}
nonwife income, age, schooling \\ 
\# kids w/ age 0-5,\ \# kids w/ age 6-18,%
\end{tabular}
\\ \cline{2-3}
& $X_{i}$ & 
\begin{tabular}{l}
mother and father's schooling \\ 
age, schooling, household income,\ and age\&school interaction%
\end{tabular}
\\ \hline
\end{tabular}%
$

\bigskip \bigskip 
\end{center}

The data sets were taken from NLSY79 for the 1998 round. The data set used
in this study contains 960 female workers, after eliminating the individuals
with missing values for demographic variables used in this study. In
particular, the data set excluded those with missing values for household's
total income, because we cannot compute nonwife income or nonlabor income
from the data sets. The income variables are in 1998 dollars. The following
table offers summary statistics of the variables, and also compares them
before and after the selection process.\bigskip

\begin{center}
\bigskip 

Table 6 : Summary Statistics of Variables\bigskip

\begin{tabular}{c|lccc|cccc}
\hline\hline
&  & {\small Whole} & {\small Sample} & {\small (}$n=1268${\small )} &  & 
{\small Selected} & {\small Sample} & {\small (}$n=960${\small )} \\ 
& {\small mean} & {\small std dev} & {\small min} & {\small max} & {\small %
mean} & {\small std dev} & {\small min} & {\small max} \\ \hline\hline
{\small mother's schooling} & {\small 11.94} & {\small 2.41} & {\small 0} & 
{\small 20} & {\small 11.95} & {\small 2.35} & {\small 0} & {\small 19} \\ 
{\small father's schooling} & {\small 12.25} & {\small 3.13} & {\small 0} & 
{\small 20} & {\small 12.29} & {\small 3.11} & {\small 1} & {\small 20} \\ 
{\small wife's labor incme} & {\small 24066.3} & {\small 20086.7} & {\small %
220} & {\small 147970} & {\small 24519.8} & {\small 19927.9} & {\small 220}
& {\small 147970} \\ 
{\small hsbd's\ labor incme} & {\small 50821.9} & {\small 40895.6} & {\small %
52} & {\small 212480} & {\small 49236.1} & {\small 38190.5} & {\small 52} & 
{\small 212480} \\ 
{\small hshld's\ total incme} & {\small 69603.0} & {\small 45415.0} & 
{\small 10} & {\small 244343} & {\small 70986.7} & {\small 45085.1} & 
{\small 700} & {\small 244343} \\ 
{\small employment status} & {\small 0.74} & {\small 0.44} & {\small 0} & 
{\small 1} & {\small 0.75} & {\small 0.43} & {\small 0} & {\small 1} \\ 
{\small wife's schooling} & {\small 13.8} & {\small 2.39} & {\small 6} & 
{\small 20} & {\small 13.9} & {\small 2.31} & {\small 7} & {\small 20} \\ 
{\small wife's age} & {\small 36.9} & {\small 2.23} & {\small 33} & {\small %
41} & {\small 36.9} & {\small 2.22} & {\small 33} & {\small 41} \\ 
{\small wife's hours} & {\small 1447.4} & {\small 966.1} & {\small 0} & 
{\small 6708} & {\small 1.464.1} & {\small 948.1} & {\small 0} & {\small 5200%
} \\ 
{\small husband's age} & {\small 39.4} & {\small 5.09} & {\small 26} & 
{\small 70} & {\small 39.24} & {\small 4.92} & {\small 27} & {\small 62} \\ 
{\small husband's schooling} & {\small 13.8} & {\small 2.61} & {\small 3} & 
{\small 20} & {\small 13.7} & {\small 2.50} & {\small 3} & {\small 20} \\ 
{\small \# kids w/ age 0-5} & {\small 0.46} & {\small 0.73} & {\small 0} & 
{\small 4} & {\small 0.47} & {\small 0.73} & {\small 0} & {\small 4} \\ 
{\small \# kids w/ age 6-18} & {\small 1.33} & {\small 1.11} & {\small 0} & 
{\small 6} & {\small 1.32} & {\small 1.08} & {\small 0} & {\small 5} \\ 
\hline
\end{tabular}

\bigskip \bigskip 
\end{center}

In this study, we focus on how the estimates of coefficients in the outcome
equation vary across different specifications of $X_{i}$ and different
methods of estimating $\theta _{0}$ in the participation equation. We
estimated the model using three different estimation methods. The first
method is OLS, ignoring sample selection. The second method is Heckman's two
step approach assuming the joint normality of the errors in the outcome and
selection equations. The third method employs a semiparametric approach
through the formulation of partial linear model and following the procedure
of Robinson (1988). As for the third method, this study considered two
different methods of estimating the coefficients to $X_{i}:$ probit and
maximum score estimation.

The results are shown in Tables 6-9. (The covariates in $Z_{i}$ and $X_{i}$
were appropriately rescaled to ensure numerical stability.) First, nonwife
income and the number of young and old children play a significant role in
determining the labor supply of female workers. This result is robust
through different model specifications, although the significance of nonwife
income is somewhat reduced when one incorporates sample selection. The
negative effect of the number of children is conspicuous, with the effect of
the number of young children stronger than that of old children. In
contrast, the significant role that the female worker's age and schooling
appear to play in the case of OLS or Heckman's two step procedure with the
joint normality assumption disappears or is substantially reduced, when one
moves to a semiparametric model.

Finally, it is interesting to observe that within the framework of a
semiparametric model, the effects of log wage and nonwife income on labor
participation are shown to be more significant in the case of using maximum
score estimation than in the case of using a probit estimator $\hat{\theta}$%
. This appears to be some evidence against the assumption that $X_{i}$ and $%
\varepsilon _{i}$ in the selection equation are independent. While a formal
testing procedure seems appropriate, a direct test comparing the estimates
are not available in the literature as far as the author is concerned. In
particular, the standard Hausman type test will not have an asymptotically
exact size because when the probit estimator and the maximum score estimator
are both consistent, the asymptotic distribution of $\sqrt{n}\{\hat{\beta}(%
\hat{\theta}_{probit})-\hat{\beta}(\hat{\theta}_{mx.scr})\}$, $\hat{\theta}%
_{probit}$ denoting a probit estimator of $\theta _{0}$ and $\hat{\theta}%
_{mx.scr}$ a maximum score estimator, will be degenerate. The latter
degeneracy is a major implication of Theorem 1 in this paper.

\begin{center}
\bigskip 

Table 7: Estimation of Female Labor Participation (Specification I)

(In the parentheses are standard errors.)\bigskip

\begin{tabular}{c|r|r|cl}
\hline\hline
& {\small \ \ \ \ \ \ \ \ \ OLS\ \ \ \ \ \ \ \ \ } & {\small \ \ \ Mill's
Ratio\ \ \ } & {\small Semiparametric} & {\small Model\ \ \ \ \ \ \ \ \ \ \
\ \ \ \ } \\ 
& \multicolumn{1}{|c|}{} & \multicolumn{1}{|c|}{} & $\hat{\theta}${\small \
w/ Probit} & \multicolumn{1}{c}{$\hat{\theta}${\small \ w/Mx.\ Scr}} \\ 
\hline\hline
{\small Log Wage} & \multicolumn{1}{|c|}{${\small -37.500}$} & 
\multicolumn{1}{|c|}{${\small -41.857}$} & ${\small -46.764}$ & 
\multicolumn{1}{c}{${\small -50.458}$} \\ 
& \multicolumn{1}{|c|}{${\small (40.890)}$} & \multicolumn{1}{|c|}{${\small %
(21.520)}$} & ${\small (60.550)}$ & \multicolumn{1}{c}{${\small (60.799)}$}
\\ \hline
{\small Nonwife Income} & \multicolumn{1}{|c|}{${\small -0.0248}$} & 
\multicolumn{1}{|c|}{${\small -0.0251}$} & ${\small -0.0197}$ & 
\multicolumn{1}{c}{${\small -0.0217}$} \\ 
& \multicolumn{1}{|c|}{${\small (0.0080)}$} & \multicolumn{1}{|c|}{${\small %
(0.0042)}$} & ${\small (0.0111)}$ & \multicolumn{1}{c}{${\small (0.0111)}$}
\\ \hline
{\small Young Children} & \multicolumn{1}{|c|}{${\small -0.1491}$} & 
\multicolumn{1}{|c|}{${\small -0.1558}$} & ${\small -0.1748}$ & 
\multicolumn{1}{c}{${\small -0.1761}$} \\ 
& \multicolumn{1}{|c|}{${\small (0.0424)}$} & \multicolumn{1}{|c|}{${\small %
(0.0223)}$} & ${\small (0.0390)}$ & \multicolumn{1}{c}{${\small (0.0395)}$}
\\ \hline
{\small Old Children} & \multicolumn{1}{|c|}{${\small -0.1217}$} & 
\multicolumn{1}{|c|}{${\small -0.1233}$} & ${\small -0.1310}$ & 
\multicolumn{1}{c}{${\small -0.1305}$} \\ 
& \multicolumn{1}{|c|}{${\small (0.0258)}$} & \multicolumn{1}{|c|}{${\small %
(0.0137)}$} & ${\small (0.0251)}$ & \multicolumn{1}{c}{${\small (0.0249)}$}
\\ \hline
{\small Age} & \multicolumn{1}{|c|}{${\small 0.0466}$} & 
\multicolumn{1}{|c|}{${\small 0.0438}$} & ${\small -0.0024}$ & 
\multicolumn{1}{c}{${\small 0.0097}$} \\ 
& \multicolumn{1}{|c|}{${\small (0.0046)}$} & \multicolumn{1}{|c|}{${\small %
(0.0030)}$} & ${\small (0.0131)}$ & \multicolumn{1}{c}{${\small (0.0163)}$}
\\ \hline
{\small Schooling} & \multicolumn{1}{|c|}{${\small 0.0368}$} & 
\multicolumn{1}{|c|}{${\small 0.0382}$} & ${\small 0.0158}$ & 
\multicolumn{1}{c}{${\small 0.0227}$} \\ 
& \multicolumn{1}{|c|}{${\small (0.0120)}$} & \multicolumn{1}{|c|}{${\small %
(0.0078)}$} & ${\small (0.0145)}$ & \multicolumn{1}{c}{${\small (0.0228)}$}
\\ \hline
\end{tabular}

\bigskip 

Table 8: Estimation of Female Labor Participation (Specification II)

(In the parentheses are standard errors.)\bigskip

\begin{tabular}{c|r|r|cl}
\hline\hline
& {\small \ \ \ \ \ \ \ \ \ OLS\ \ \ \ \ \ \ \ \ } & {\small \ \ \ Mill's
Ratio\ \ \ } & {\small Semiparametric} & {\small Model\ \ \ \ \ \ \ \ \ \ \
\ \ \ \ } \\ 
& \multicolumn{1}{|c|}{} & \multicolumn{1}{|c|}{} & $\hat{\theta}${\small \
w/ Probit} & \multicolumn{1}{c}{$\hat{\theta}${\small \ w/Mx.\ Scr}} \\ 
\hline\hline
{\small Log Wage} & \multicolumn{1}{|c|}{${\small -37.500}$} & 
\multicolumn{1}{|c|}{${\small -37.800}$} & ${\small -50.691}$ & 
\multicolumn{1}{c}{${\small -212.93}$} \\ 
& \multicolumn{1}{|c|}{${\small (40.890)}$} & \multicolumn{1}{|c|}{${\small %
(2.9290)}$} & ${\small (59.971)}$ & \multicolumn{1}{c}{${\small (65.328)}$}
\\ \hline
{\small Nonwife Income} & \multicolumn{1}{|c|}{${\small -0.0248}$} & 
\multicolumn{1}{|c|}{${\small -0.0247}$} & ${\small -0.0222}$ & 
\multicolumn{1}{c}{${\small -0.0937}$} \\ 
& \multicolumn{1}{|c|}{${\small (0.0080)}$} & \multicolumn{1}{|c|}{${\small %
(0.0007)}$} & ${\small (0.0110)}$ & \multicolumn{1}{c}{${\small (0.0172)}$}
\\ \hline
{\small Young Children} & \multicolumn{1}{|c|}{${\small -0.1491}$} & 
\multicolumn{1}{|c|}{${\small -0.1500}$} & ${\small -0.1724}$ & 
\multicolumn{1}{c}{${\small -0.1543}$} \\ 
& \multicolumn{1}{|c|}{${\small (0.0424)}$} & \multicolumn{1}{|c|}{${\small %
(0.0030)}$} & ${\small (0.0385)}$ & \multicolumn{1}{c}{${\small (0.0374)}$}
\\ \hline
{\small Old Children} & \multicolumn{1}{|c|}{${\small -0.1217}$} & 
\multicolumn{1}{|c|}{${\small -0.1219}$} & ${\small -0.1305}$ & 
\multicolumn{1}{c}{${\small -0.1140}$} \\ 
& \multicolumn{1}{|c|}{${\small (0.0258)}$} & \multicolumn{1}{|c|}{${\small %
(0.0018)}$} & ${\small (0.0252)}$ & \multicolumn{1}{c}{${\small (0.0234)}$}
\\ \hline
{\small Age} & \multicolumn{1}{|c|}{${\small 0.0466}$} & 
\multicolumn{1}{|c|}{${\small 0.0462}$} & ${\small -0.0097}$ & 
\multicolumn{1}{c}{${\small 0.0199}$} \\ 
& \multicolumn{1}{|c|}{${\small (0.0046)}$} & \multicolumn{1}{|c|}{${\small %
(0.0004)}$} & ${\small (0.0131)}$ & \multicolumn{1}{c}{${\small (0.0123)}$}
\\ \hline
{\small Schooling} & \multicolumn{1}{|c|}{${\small 0.0368}$} & 
\multicolumn{1}{|c|}{${\small 0.0369}$} & ${\small 0.0124}$ & 
\multicolumn{1}{c}{${\small -0.0226}$} \\ 
& \multicolumn{1}{|c|}{${\small (0.0120)}$} & \multicolumn{1}{|c|}{${\small %
(0.0010)}$} & ${\small (0.0143)}$ & \multicolumn{1}{c}{${\small (0.0132)}$}
\\ \hline
\end{tabular}

\bigskip \bigskip \pagebreak 

\bigskip 

Table 9: Estimation of Female Labor Participation (Specification III)

(In the parentheses are standard errors.)\bigskip

\begin{tabular}{c|r|r|cl}
\hline\hline
& {\small \ \ \ \ \ \ \ \ \ OLS\ \ \ \ \ \ \ \ \ } & {\small \ \ \ Mill's
Ratio\ \ \ } & {\small Semiparametric} & {\small Model\ \ \ \ \ \ \ \ \ \ \
\ \ \ \ } \\ 
& \multicolumn{1}{|c|}{} & \multicolumn{1}{|c|}{} & $\hat{\theta}${\small \
w/ Probit} & \multicolumn{1}{c}{$\hat{\theta}${\small \ w/Mx.\ Scr}} \\ 
\hline\hline
{\small Log Wage} & \multicolumn{1}{|c|}{${\small -37.500}$} & 
\multicolumn{1}{|c|}{${\small -49.653}$} & ${\small -48.206}$ & 
\multicolumn{1}{c}{${\small -145.471}$} \\ 
& \multicolumn{1}{|c|}{${\small (40.890)}$} & \multicolumn{1}{|c|}{${\small %
(83.915)}$} & ${\small (59.261)}$ & \multicolumn{1}{c}{${\small (61.158)}$}
\\ \hline
{\small Nonwife Income} & \multicolumn{1}{|c|}{${\small -0.0248}$} & 
\multicolumn{1}{|c|}{${\small -0.0218}$} & ${\small -0.0226}$ & 
\multicolumn{1}{c}{${\small -0.0689}$} \\ 
& \multicolumn{1}{|c|}{${\small (0.0080)}$} & \multicolumn{1}{|c|}{${\small %
(0.0245)}$} & ${\small (0.0110)}$ & \multicolumn{1}{c}{${\small (0.0137)}$}
\\ \hline
{\small Young Children} & \multicolumn{1}{|c|}{${\small -0.1491}$} & 
\multicolumn{1}{|c|}{${\small -0.1766}$} & ${\small -0.1735}$ & 
\multicolumn{1}{c}{${\small -0.1557}$} \\ 
& \multicolumn{1}{|c|}{${\small (0.0424)}$} & \multicolumn{1}{|c|}{${\small %
(0.0809)}$} & ${\small (0.0382)}$ & \multicolumn{1}{c}{${\small (0.0366)}$}
\\ \hline
{\small Old Children} & \multicolumn{1}{|c|}{${\small -0.1217}$} & 
\multicolumn{1}{|c|}{${\small -0.1301}$} & ${\small -0.1291}$ & 
\multicolumn{1}{c}{${\small -0.1176}$} \\ 
& \multicolumn{1}{|c|}{${\small (0.0258)}$} & \multicolumn{1}{|c|}{${\small %
(0.0477)}$} & ${\small (0.0250)}$ & \multicolumn{1}{c}{${\small (0.0241)}$}
\\ \hline
{\small Age} & \multicolumn{1}{|c|}{${\small 0.0466}$} & 
\multicolumn{1}{|c|}{${\small 0.0079}$} & ${\small -0.0056}$ & 
\multicolumn{1}{c}{${\small 0.0028}$} \\ 
& \multicolumn{1}{|c|}{${\small (0.0046)}$} & \multicolumn{1}{|c|}{${\small %
(0.0142)}$} & ${\small (0.0122)}$ & \multicolumn{1}{c}{${\small (0.0120)}$}
\\ \hline
{\small Schooling} & \multicolumn{1}{|c|}{${\small 0.0368}$} & 
\multicolumn{1}{|c|}{${\small 0.0244}$} & ${\small 0.0163}$ & 
\multicolumn{1}{c}{${\small -0.0112}$} \\ 
& \multicolumn{1}{|c|}{${\small (0.0120)}$} & \multicolumn{1}{|c|}{${\small %
(0.0387)}$} & ${\small (0.0144)}$ & \multicolumn{1}{c}{${\small (0.0133)}$}
\\ \hline
\end{tabular}

\bigskip \bigskip 
\end{center}

\section{Conclusion}

This paper focuses on semiparametric models where the identified parameter
involves conditional expectations with a single-index as a conditioning
variable. This paper offers a set of sufficient conditions under which the
first step estimator of the single-index does not have a first order impact
on the asymptotic distribution of the second step estimator. The remarkable
aspect of the result is that the asymptotic negligibility of the first step
estimator holds even when the estimator follows cube-root asymptotics. This
asymptotic negligibility is also demonstrated through Monte Carlo simulation
studies. The usefulness of this procedure is illustrated by an empirical
study of female labor supply using an NLSY79 data set.

\section{The Appendix}

\subsection{Proof of Theorem 1}

\noindent \textsc{Lemma A1:} \textit{Suppose that Assumptions C1, G1(i)(ii),
and G2(i) hold. Then, there exist }$C>0$\textit{\ and }$\varepsilon >0$ 
\textit{such that for each }$\eta \in (0,\varepsilon ],$%
\begin{equation*}
\sup_{\theta \in \mathbf{R}^{d}:||\theta -\theta _{0}||\leq \eta }\left\Vert
a(\theta )-a(\theta _{0})\right\Vert \leq C\eta ^{2}.
\end{equation*}

\noindent \textsc{Proof:} The result follows from Theorem 1 of Song (2012).
See Example 1 there. $\blacksquare $\bigskip

\noindent \textsc{Proof of Theorem 1:} Without loss of generality, let $K$
be $[-1/2,1/2]$-supported. It suffices to focus on the case where $H$ is $%
\mathbf{R}$-valued. Let $\hat{\mu}_{\theta }$ and $\hat{U}_{\theta ,i}\ $be $%
\hat{\mu}$ and $\hat{U}_{i}$ except that $\hat{\theta}$ is replaced by $%
\theta $, and let $\mu _{\theta }$ and $U_{\theta ,i}$ be $\mu $ and $%
U_{0,i} $ except that $\theta _{0}$ is replaced by $\theta $, where $%
U_{0,i}=F_{\theta _{0}}(X_{i}^{\top }\theta _{0})$. First, observe that%
\begin{eqnarray*}
\hat{a}(\theta )-a(\theta ) &=&\frac{1}{\sum_{i=1}^{n}D_{i}}%
\sum_{i=1}^{n}D_{i}S_{i}\cdot \left\{ \varphi (\hat{\mu}_{\theta }(\hat{U}%
_{\theta ,i}))-\varphi (\mu _{\theta }(U_{\theta ,i}))\right\} \\
&&+\frac{1}{\sum_{i=1}^{n}D_{i}}\sum_{i=1}^{n}\left\{ D_{i}S_{i}\cdot
\varphi (\mu _{\theta }(U_{\theta ,i}))-\mathbf{E}\left[ D_{i}S_{i}\cdot
\varphi (\mu _{\theta }(U_{\theta ,i}))\right] \right\} \\
&&+\left\{ \frac{1}{\frac{1}{n}\sum_{i=1}^{n}D_{i}}-\frac{1}{P\{D_{i}=1\}}%
\right\} \mathbf{E}\left[ D_{i}S_{i}\cdot \varphi (\mu _{\theta }(U_{\theta
,i}))\right] \\
&\equiv &A_{1n}(\theta )+A_{2n}(\theta )+A_{3n}(\theta ),\text{ say.}
\end{eqnarray*}%
We write $A_{1n}(\theta )$ as%
\begin{eqnarray*}
&&\frac{1}{\sum_{i=1}^{n}D_{i}}\sum_{i=1}^{n}D_{i}S_{i}\cdot \varphi
^{\prime }(\mu _{\theta }(U_{\theta ,i}))\left\{ \hat{\mu}_{\theta }(\hat{U}%
_{\theta ,i})-\mu _{\theta }(U_{\theta ,i})\right\} \\
&&+\frac{1}{2\sum_{i=1}^{n}D_{i}}\sum_{i=1}^{n}D_{i}S_{i}\cdot \varphi
^{\prime \prime }(\mu _{\theta }^{\ast }(U_{\theta ,i}))\left\{ \hat{\mu}%
_{\theta }(\hat{U}_{\theta ,i})-\mu _{\theta }(U_{\theta ,i})\right\} ^{2} \\
&\equiv &B_{1n}(\theta )+B_{2n}(\theta )\text{, say,}
\end{eqnarray*}%
where $\mu _{\theta }^{\ast }(U_{\theta ,i})$ lies on the line segment
between $\hat{\mu}_{\theta }(\hat{U}_{\theta ,i})$ and $\mu _{\theta
}(U_{\theta ,i})$. Let $1_{n,i}=1\{|U_{0,i}-1|>h/2\}$, and write $%
B_{2n}(\theta )$ as%
\begin{eqnarray*}
&&\frac{1}{2\sum_{i=1}^{n}D_{i}}\sum_{i=1}^{n}D_{i}S_{i}\cdot \varphi
^{\prime \prime }(\mu _{\theta }^{\ast }(U_{\theta ,i}))\left\{ \hat{\mu}%
_{\theta }(\hat{U}_{\theta ,i})-\mu _{\theta }(U_{\theta ,i})\right\}
^{2}1_{n,i} \\
&&+\frac{1}{2\sum_{i=1}^{n}D_{i}}\sum_{i=1}^{n}D_{i}S_{i}\cdot \varphi
^{\prime \prime }(\mu _{\theta }^{\ast }(U_{\theta ,i}))\left\{ \hat{\mu}%
_{\theta }(\hat{U}_{\theta ,i})-\mu _{\theta }(U_{\theta ,i})\right\}
^{2}\left\{ 1-1_{n,i}\right\} \\
&\equiv &C_{1n}(\theta )+C_{2n}(\theta )\text{, say.}
\end{eqnarray*}%
For small $\varepsilon >0$ and a positive sequence $c_{n}>0$ such that $%
0<c_{n}n^{1/4}\rightarrow 0$ (see, e.g., the proof of Lemma A3 of Song
(2009)),%
\begin{eqnarray}
\max_{1\leq i\leq n}\sup_{\theta \in B(\theta _{0};\varepsilon )}\left\vert 
\hat{U}_{\theta ,i}-U_{\theta ,i}\right\vert &=&O_{P}(1/\sqrt{n})\text{ and}
\label{cvU} \\
\max_{1\leq i\leq n}\sup_{\theta \in B(\theta _{0};c_{n})}\left\vert
U_{\theta ,i}-U_{0,i}\right\vert &=&O_{P}(c_{n}).  \notag
\end{eqnarray}%
By Assumption C3(ii), $c_{n}h^{-1}\rightarrow 0$. As for $C_{1n}(\theta )$,
we bound $\left\vert \hat{\mu}_{\theta }(\hat{U}_{\theta ,i})-\mu _{\theta
}(U_{\theta ,i})\right\vert 1_{n,i}$ by (using (\ref{cvU}))%
\begin{equation*}
\sup_{u\in \lbrack h/2,1-h/2]}\left\vert \hat{\mu}_{\theta }(u)-\mu _{\theta
}(u)\right\vert +\left\vert \mu _{\theta }(\hat{U}_{\theta ,i})-\mu _{\theta
}(U_{\theta ,i})\right\vert 1_{n,i}
\end{equation*}%
from some large $n$ on. The first term is $O(n^{-1/2}h^{-1}\sqrt{\log n})$
(e.g. see Lemma A4 of Song (2009)), and the second term is $O_{P}(1/\sqrt{n}%
),$ both uniformly over $1\leq i\leq n$ and over $\theta \in B(\theta
_{0};c_{n})$. The latter rate $O_{P}(1/\sqrt{n})$ stems from (\ref{cvU}),
and that $\mu _{\theta }(\cdot )$ is continuously differentiable with a
bounded derivative (see Assumption G2(ii)). Since $\varphi ^{\prime \prime
}(\cdot )$ is continuous and bounded on the support of $\mathbf{E}[W|X,D=1]$
(Assumptions C1(i) and G1(ii)(a)), $\sup_{\theta \in B(\theta
_{0};\varepsilon )}|C_{1n}(\theta )|=O_{P}(n^{-1}h^{-2}(\log n))$. As for $%
C_{2n}(\theta )$, we bound sup$_{\theta \in B(\theta _{0};c_{n})}|\hat{\mu}%
_{\theta }(\hat{U}_{\theta ,i})-\mu _{\theta }(U_{\theta
,i})|(1-1_{n,i})=O_{P}(h+n^{-1/2})$, uniformly over $1\leq i\leq n$. The
rate $O_{P}(h)$ here is the convergence rate at the boundary (e.g. see Lemma
A4 of Song (2009)). Therefore, for some $C>0$,%
\begin{eqnarray*}
\sup_{\theta \in B(\theta _{0};c_{n})}|C_{2n}(\theta )| &\leq &C\frac{1}{n}%
\sum_{i=1}^{n}\left\{ \hat{\mu}_{\theta }(\hat{U}_{\theta ,i})-\mu _{\theta
}(U_{\theta ,i})\right\} ^{2}\left\vert 1-1_{n,i}\right\vert \\
&\leq &C\max_{1\leq j\leq n}\sup_{\theta \in B(\theta _{0};c_{n})}\left\{ 
\hat{\mu}_{\theta }(\hat{U}_{\theta ,j})-\mu _{\theta }(U_{\theta
,j})\right\} ^{2}\left\vert 1-1_{n,j}\right\vert \\
&&\cdot \frac{1}{n}\sum_{i=1}^{n}\left\vert 1-1_{n,i}\right\vert \\
&=&O_{P}(h^{2}+n^{-1})\cdot O_{P}(h)=o_{P}(n^{-1/2}).
\end{eqnarray*}%
We conclude $\sup_{\theta \in B(\theta _{0};c_{n})}|A_{1n}(\theta
)-B_{1n}(\theta )|=o_{P}\left( n^{-1/2}\right) .$ As for $B_{1n}(\theta )$,
we apply Lemma B3 below to deduce that with $P_{1}\equiv P\{D_{i}=1\}$,%
\begin{equation*}
\frac{1}{P_{1}}\cdot \frac{1}{n}\sum_{i=1}^{n}D_{i}\mathbf{E}\left[
S_{i}\cdot \varphi ^{\prime }(\mu _{\theta }(U_{\theta ,i}))|U_{\theta
,i},D_{i}=1\right] \left\{ \varphi (W_{i})-\mu _{\theta }(U_{\theta
,i})\right\} +o_{P}(1/\sqrt{n})\text{,}
\end{equation*}%
uniformly over $\theta \in B(\theta _{0};c_{n})$. (With the help of Lemma B1
below, one can check that Assumptions B1-B3 in Lemma B3 are satisfied by the
conditions of the theorem here.)

We turn to $A_{3n}(\theta )$, which we write as%
\begin{equation*}
\mathbf{E}\left[ S_{i}\cdot \varphi (\mu _{\theta }(U_{\theta ,i}))|D_{i}=1%
\right] \cdot \frac{1}{P_{1}n}\sum_{i=1}^{n}\{P\{D=1\}-D_{i}\}+o_{P}(1/\sqrt{%
n})
\end{equation*}%
uniformly over $\theta \in B(\theta _{0};c_{n})$. Combining the results so
far, we find that $\sqrt{n}\{\hat{a}(\theta )-a(\theta )\}$ is equal to%
\begin{eqnarray*}
&&\frac{1}{P_{1}\sqrt{n}}\sum_{i=1}^{n}D_{i}\mathbf{E}\left[ S_{i}\cdot
\varphi ^{\prime }(\mu _{\theta }(U_{\theta ,i}))|U_{\theta ,i},D_{i}=1%
\right] \left\{ \varphi (W_{i})-\mu _{\theta }(U_{\theta ,i})\right\} \\
&&+\frac{1}{P_{1}\sqrt{n}}\sum_{i=1}^{n}\left\{ D_{i}S_{i}\cdot \varphi (\mu
_{\theta }(U_{\theta ,i}))-\mathbf{E}\left[ D_{i}S_{i}\cdot \varphi (\mu
_{\theta }(U_{\theta ,i}))\right] \right\} \\
&&+\mathbf{E}\left[ S_{i}\cdot \varphi (\mu _{\theta }(U_{\theta
,i}))|D_{i}=1\right] \cdot \frac{1}{P_{1}\sqrt{n}}\sum_{i=1}^{n}\{P\left\{
D=1\right\} -D_{i}\}+o_{P}(1),
\end{eqnarray*}%
uniformly over $\theta \in B(\theta _{0};c_{n})$. From this uniform linear
representation of $\sqrt{n}\{\hat{a}(\theta )-a(\theta )\},$ it is not hard
to show that 
\begin{eqnarray}
\sup_{\theta \in B(\theta _{0};c_{n})}|\sqrt{n}\{\hat{a}(\theta )-a(\theta
)\}| &=&O_{P}(1)\text{ and}  \label{conv7} \\
\sup_{\theta \in B(\theta _{0};c_{n})}\left\vert \sqrt{n}\{\hat{a}(\theta
)-a(\theta )-(\hat{a}(\theta _{0})-a(\theta _{0}))\}\right\vert &=&o_{P}(1).
\notag
\end{eqnarray}%
(For this, we can use Lemma B1 below to obtain an entropy bound for the
class of functions indexing the processes in the linear representation of $%
\sqrt{n}\{\hat{a}(\theta )-a(\theta )\}$. Details are omitted.)

Let $H_{1}(a,b)=\partial H(a,b)/\partial a$, and let the sequence $c_{n}$
chosen above be such that $||\hat{\theta}-\theta _{0}||=O_{P}(c_{n})$ and $%
\sqrt{n}c_{n}^{2}\rightarrow 0$ as $n\rightarrow \infty $. We write%
\begin{eqnarray*}
\sqrt{n}(\hat{\beta}-\tilde{\beta}) &=&\sqrt{n}\{H(\hat{a}(\hat{\theta}),%
\hat{b})-H(\hat{a}(\theta _{0}),\hat{b})\} \\
&=&H_{1}(a(\theta _{0}),b_{0})^{\top }\sqrt{n}\{\hat{a}(\hat{\theta})-\hat{a}%
(\theta _{0})\}+o_{P}(1) \\
&=&H_{1}(a(\theta _{0}),b_{0})^{\top }\sqrt{n}\{\hat{a}(\hat{\theta})-a(\hat{%
\theta})-\hat{a}(\theta _{0})+a(\theta _{0})\} \\
&&+H_{1}(a(\theta _{0}),b_{0})^{\top }\sqrt{n}\{a(\hat{\theta})-a(\theta
_{0})\}+o_{P}(1)\equiv D_{1n}+D_{2n}\text{, say.}
\end{eqnarray*}%
The second equality uses the first statement of (\ref{conv7}) and continuity
of $H_{1}(\cdot ,\cdot )$ (Assumption G1(ii)(b)). By the second statement of
(\ref{conv7}), $D_{1n}=o_{P}(1)$. As for $D_{2n}$, we apply Lemma A1 to
deduce that%
\begin{equation*}
|H_{1}(a(\theta _{0}),b_{0})^{\top }\sqrt{n}\{a(\hat{\theta})-a(\theta
_{0})\}|=O_{P}(n^{1/2}\times c_{n}^{2})=o_{P}(1).
\end{equation*}%
Thus we obtain the desired result. $\blacksquare $

\subsection{Bahadur Representation of Sample Linear Functionals of SNN
Estimators}

In this section, we present a Bahadur representation of sample linear
functionals of SNN estimators that is uniform over function spaces. In a
different context, Stute and Zhu (2005) obtained a related result that is
not uniform.

Suppose that we are given a random sample $\{(S_{i},W_{i},X_{i})\}_{i=1}^{n}$
drawn from the distribution of a random vector $(S,W,X)\in \mathbf{R}%
^{d_{S}+1+d_{X}}.$\ Let $\mathcal{S}_{S},\mathcal{S}_{X}$ and $\mathcal{S}%
_{W}$ be the supports of $S,X,$ and $W$ respectively. Let $\Lambda \ $be a
class of $\mathbf{R}$-valued functions on $\mathbf{R}^{d_{X}}$ with a
generic element denoted by $\lambda .$ We also let $\Phi $ and $\Psi $ be
classes of real functions on $\mathbf{R}$ and $\mathbf{R}^{d_{S}}\ $with
generic elements $\varphi $ and $\psi $ and let $\tilde{\varphi}$ and\ $%
\tilde{\psi}$ be their envelopes. Let $L_{p}(P),\ p\geq 1,$ be the space of $%
L_{p}$-bounded functions: $||f||_{p}\equiv \{\int
|f(x)|^{p}P(dx)\}^{1/p}<\infty ,$ and for a space of functions $\mathcal{F}%
\subset L_{p}(P)$ for $p\geq 1,\ $let $N_{[]}(\varepsilon ,\mathcal{F}%
,||\cdot ||_{p})$ denote the bracketing number of $\mathcal{F}$ with respect
to the norm $||\cdot ||_{p}$, i.e., the smallest number $r$ such that there
exist $f_{1},\cdot \cdot \cdot ,f_{r}$ and $\Delta _{1},\cdot \cdot \cdot
,\Delta _{r}\in L_{p}(P)$ such that $||\Delta _{i}||_{p}<\varepsilon $ and
for all $f\in \mathcal{F}$, there exists $1\leq i\leq r$ with $%
||f_{i}-f||_{p}<\Delta _{i}/2.$ Similarly, we define $N_{[]}(\varepsilon ,%
\mathcal{F},||\cdot ||_{\infty })$ to be the bracketing number of $\mathcal{F%
}$ with respect to the sup norm $||\cdot ||_{\infty }$, where for any real
map $f$ on $\mathbf{R}^{d_{X}},$ we define $||f||_{\infty }=\sup_{z\in 
\mathbf{R}^{d_{X}}}|f(z)|$. For any norm $||\cdot ||$ which is equal to $%
||\cdot ||_{p}$ or $||\cdot ||_{\infty }$, we define $N(\varepsilon ,%
\mathcal{F},||\cdot ||)$ to be the covering number of $\mathcal{F}$, i.e.,
the smallest number of $\varepsilon $-balls that cover $\mathcal{F}$.
Letting $F_{\lambda }(\cdot )$ be the CDF of $\lambda (X)$, we denote $%
U_{\lambda }\equiv F_{\lambda }(\lambda (X)).$ Define $g_{\varphi ,\lambda
}(u)\equiv \mathbf{E}[\varphi (W)|U_{\lambda }=u]\ $and $g_{\psi ,\lambda
}(u)\equiv \mathbf{E}[\psi (S)|U_{\lambda }=u].$

Let $U_{n,\lambda ,i}\equiv \frac{1}{n-1}\sum_{j=1,j\neq i}^{n}1\{\lambda
(X_{j})\leq \lambda (X_{i})\}$ and consider the estimator:%
\begin{equation*}
\hat{g}_{\varphi ,\lambda ,i}(u)\equiv \frac{1}{(n-1)\hat{f}_{\lambda ,i}(u)}%
\sum_{j=1,j\neq i}^{n}\varphi (W_{j})K_{h}\left( U_{n,\lambda ,j}-u\right) ,
\end{equation*}%
where $\hat{f}_{\lambda ,i}(u)\equiv (n-1)^{-1}\sum_{j=1,j\neq
i}^{n}K_{h}(U_{n,\lambda ,j}-u).$ The semiparametric process of focus takes
the following form:%
\begin{equation*}
\nu _{n}(\lambda ,\varphi ,\psi )\equiv \frac{1}{\sqrt{n}}\sum_{i=1}^{n}\psi
(S_{i})\left\{ \hat{g}_{\varphi ,\lambda ,i}(U_{n,\lambda ,i})-g_{\varphi
}(U_{\lambda ,i})\right\} ,
\end{equation*}%
with $(\lambda ,\varphi ,\psi )\in \Lambda \times \Phi \times \Psi .$

The main focus in this section is on establishing an asymptotic linear
representation of $\nu _{n}(\lambda ,\varphi ,\psi )$. The critical element
in the proof is to bound the size of the class of conditional expectation
functions $\mathcal{G}\equiv \{g_{\varphi ,\lambda }(\cdot ):(\varphi
,\lambda )\in \Phi \times \Lambda \}$. We begin with the following lemma
that establishes the bracketing entropy bound for $\mathcal{G}$ with respect
to $||\cdot ||_{q},\ q\geq 1$.\bigskip

\noindent \textsc{Lemma B1} \textsc{:}\textbf{\ }\textit{Suppose that the
density }$f_{\lambda }$ \textit{of }$\lambda (X)$\textit{\ is bounded
uniformly over }$\lambda \in \Lambda $. \textit{Furthermore, assume that
there exists an envelope }$\tilde{\varphi}$ \textit{for }$\Phi $\textit{\
such that }$G_{\Phi }\equiv \sup_{x\in \mathcal{S}_{X}}\mathbf{E}[\tilde{%
\varphi}(W)|X=x]<\infty ,$ \textit{and that for some }$C_{L}>0,$%
\begin{equation*}
\sup_{\varphi \in \Phi }\sup_{\lambda \in \Lambda }\left\vert g_{\varphi
,\lambda }(u_{1})-g_{\varphi ,\lambda }(u_{2})\right\vert \leq
C_{L}|u_{1}-u_{2}|\text{, \textit{for all} }u_{1},u_{2}\in \lbrack 0,1].
\end{equation*}%
\textit{Then for all }$\varepsilon >0,$ $q\geq 1,$\textit{\ and }$p\geq 1$%
\textit{,}%
\begin{equation*}
N_{[]}(C_{\Phi }\varepsilon ^{1/(q+1)},\mathcal{G},||\cdot ||_{q})\leq
N_{[]}(\varepsilon ,\Phi ,||\cdot ||_{p})\cdot N_{[]}(\varepsilon ,\Lambda
,||\cdot ||_{\infty }),
\end{equation*}%
\textit{where }$C_{\Phi }\equiv 1+8C_{\Lambda }G_{\Phi }+C_{L}+G_{\Phi }/2$%
\textit{\ and }$C_{\Lambda }\equiv \sup_{\lambda \in \Lambda }\sup_{v\in 
\mathbf{R}}f_{\lambda }(v)$.\bigskip 

\noindent \textsc{Proof of Lemma B1} \textsc{:}\textbf{\ }For each $\eta \in
(0,1]$, we define $g_{\varphi ,\lambda ,\eta }(u)=\mathbf{E}\left[ \varphi
(W)Q_{\eta }(U_{\lambda }-u)\right] $, where $Q_{\eta }(u)=Q(u/\eta )/\eta $%
, and $Q(u)=1\{u\in \lbrack -1/2,1/2]\}$. For each $\eta \in (0,1]$, we let $%
\mathcal{G}_{\eta }=\{g_{\varphi ,\lambda ,\eta }(\cdot ):(\varphi ,\lambda
)\in \Phi \times \Lambda \}$. First, we show that for all $\eta \in (0,1]$,
and for all $\varepsilon >0$ and $p\geq 1$,%
\begin{equation}
N_{[]}((1+8C_{\Lambda }G_{\Phi })\cdot \varepsilon /\eta ,\mathcal{G}_{\eta
},||\cdot ||_{\infty })\leq N_{[]}(\varepsilon ,\Phi ,||\cdot ||_{p})\cdot
N_{[]}(\varepsilon ,\Lambda ,||\cdot ||_{\infty }).  \label{bound3}
\end{equation}%
Fix $\eta \in (0,1],$ $\varepsilon >0,\ $and $p\geq 1,$ and choose two sets
of brackets $\{\varphi _{j},\Delta _{\Phi ,j}\}_{j=1}^{N_{\Phi }}$ and $%
\{\lambda _{j},\Delta _{\Lambda ,j}\}_{j=1}^{N_{\Lambda }}$ that constitute $%
\varepsilon $-brackets for $\Phi $ and $\Lambda $, with respect to $||\cdot
||_{p}$ and $||\cdot ||_{\infty }$ respectively, where $N_{\Phi
}=N_{[]}(\varepsilon ,\Phi ,||\cdot ||_{p})$ and $N_{\Lambda
}=N_{[]}(\varepsilon ,\Lambda ,||\cdot ||_{\infty })$. Define $g_{jk,\eta
}(u)=\mathbf{E}\left[ \varphi _{j}(W)Q_{\eta }(U_{\lambda _{k}}-u)\right] $.
For any $g_{\varphi ,\lambda ,\eta }\in \mathcal{G}_{\eta }$, we can choose
the pairs $(\varphi _{j},\Delta _{\Phi ,j})$ and $(\lambda _{k},\Delta
_{\Lambda ,k})$ such that%
\begin{eqnarray*}
|\varphi (w)-\varphi _{j}(w)| &\leq &\Delta _{\Phi ,j}(w)\text{, for all }%
w\in \mathcal{S}_{W}\text{ and } \\
|\lambda (x)-\lambda _{k}(x)| &\leq &\Delta _{\Lambda ,k}(x),\text{ for all }%
x\in \mathcal{S}_{X}
\end{eqnarray*}%
and $||\Delta _{\Phi ,j}||_{p}\leq \varepsilon $ and $||\Delta _{\Lambda
,k}||_{\infty }\leq \varepsilon .$ Note that for $u\in \lbrack 0,1],$%
\begin{eqnarray*}
\left\vert g_{\varphi ,\lambda ,\eta }(u)-g_{jk,\eta }(u)\right\vert  &\leq &%
\mathbf{E}\left[ \Delta _{\Phi ,j}(W)\cdot \left\vert Q_{\eta }(U_{\lambda
}-u)\right\vert \right]  \\
&&+\mathbf{E}\left[ \tilde{\varphi}(W)\cdot \left\vert Q_{\eta }(U_{\lambda
}-u)-Q_{\eta }(U_{\lambda _{k}}-u)\right\vert \right] .
\end{eqnarray*}%
Certainly, $\mathbf{E}\left[ \Delta _{\Phi ,j}(W)\cdot \left\vert Q_{\eta
}(U_{\lambda }-u)\right\vert \right] \leq \mathbf{E}\left[ \Delta _{\Phi
,j}(W)\right] /\eta $. As for the second term, let $\Delta _{\lambda
,\lambda _{k}}=|U_{\lambda }-U_{\lambda _{k}}|$ and bound it by%
\begin{eqnarray*}
&&\frac{1}{\eta }G_{\Phi }\cdot P\left\{ \eta /2-\Delta _{\lambda ,\lambda
_{k}}\leq |U_{\lambda }-u|\leq \eta /2+\Delta _{\lambda ,\lambda
_{k}}\right\}  \\
&\leq &\frac{8}{\eta }C_{\Lambda }G_{\Phi }||\Delta _{\Lambda ,k}||_{\infty
},
\end{eqnarray*}%
because $\Delta _{\lambda ,\lambda _{k}}\leq 4C_{\Lambda }||\Delta _{\Lambda
,k}||_{\infty }.$ Take $\Delta _{jk,\eta }(u)\equiv \mathbf{E}\left[ \Delta
_{\Phi ,j}(W)\right] /\eta +8C_{\Lambda }G_{\Phi }||\Delta _{\Lambda
,k}||_{\infty }/\eta $ (a constant function), so that $||\Delta _{jk,\eta
}||_{\infty }\leq (1+8C_{\Lambda }G_{\Phi })\varepsilon /\eta $. Hence take $%
\{g_{jk,\eta },\Delta _{jk,\eta }\}_{j,k=1}^{N_{\Phi },N_{\Lambda }}$ to be $%
(1+8C_{\Lambda }G_{\Phi })\varepsilon /\eta $-brackets of $\mathcal{G}_{\eta
}$, affirming (\ref{bound3}).

We turn to $\mathcal{G}$. For any $(\varphi ,\lambda )\in \Phi \times
\Lambda $, we obtain that for $u\in \lbrack 0,1],$%
\begin{eqnarray*}
\left\vert g_{\varphi ,\lambda ,\eta }(u)-g_{\varphi ,\lambda
}(u)\right\vert  &\leq &\left\vert \mathbf{E}\left[ \{g_{\varphi ,\lambda
}(U_{\lambda })-g_{\varphi ,\lambda }(u)\}\cdot Q_{\eta }(U_{\lambda }-u)%
\right] \right\vert  \\
&&+\left\vert g_{\varphi ,\lambda }(u)\mathbf{E}\left[ Q_{\eta }(U_{\lambda
}-u)-1\right] \right\vert  \\
&\leq &C_{L}\eta +G_{\Phi }\cdot \left\vert \mathbf{E}\left[ Q_{\eta
}(U_{\lambda }-u)-1\right] \right\vert \leq C_{L}\eta +G_{\Phi }\cdot
b_{\eta }(u),
\end{eqnarray*}%
where $b_{\eta }(u)\equiv \frac{1}{2}\left( 1\left\{ u\in (1-\eta
/2,1]\right\} +1\left\{ u\in \lbrack 0,\eta /2)\right\} \right) $. The
second inequality follows by change of variables applied to the leading
term. The last inequality follows because for all $\eta \in (0,1],$%
\begin{eqnarray*}
\left\vert 1-\mathbf{E}Q_{\eta }(U_{\lambda }-u)\right\vert  &=&\left\vert 1-%
\frac{1}{\eta }\int_{0}^{1}1\{|v-u|\leq \eta /2\}dv\right\vert  \\
&=&\left\vert 1-\frac{1}{\eta }\int_{[u-\eta /2,u+\eta /2]\cap \lbrack
0,1]}dv\right\vert \leq b_{\eta }(u).
\end{eqnarray*}

Fix $\varepsilon >0$ and $q\geq 1$. We select $\eta =\varepsilon ^{q/(q+1)}$
and take $((1+8C_{\Lambda }G_{\Phi })\varepsilon /\eta )$-brackets $%
\{g_{jk,\eta },\Delta _{jk,\eta }\}_{j,k=1}^{N_{\Phi },N_{\Lambda }}$ that
cover $\mathcal{G}_{\eta }$ (with respect to $||\cdot ||_{\infty }$) \ with $%
N_{\Phi }=N_{[]}(\varepsilon ,\Phi ,||\cdot ||_{p})$ and $N_{\Lambda
}=N_{[]}(\varepsilon ,\Lambda ,||\cdot ||_{\infty })$. We define%
\begin{equation*}
\tilde{\Delta}_{jk,\eta }(u)\equiv \Delta _{jk,\eta }(u)+C_{L}\eta +G_{\Phi
}\cdot b_{\eta }(u).
\end{equation*}%
Then, certainly, 
\begin{eqnarray*}
||\tilde{\Delta}_{jk}||_{q} &\leq &(1+8C_{\Lambda }G_{\Phi })\varepsilon
/\eta +C_{L}\eta +G_{\Phi }\cdot ||b_{\eta }||_{q} \\
&\leq &(1+8C_{\Lambda }G_{\Phi })\varepsilon /\eta +(C_{L}+G_{\Phi }/2)\eta
^{1/q}.
\end{eqnarray*}%
Therefore, the set $\{g_{jk,\eta },\tilde{\Delta}_{jk,\eta
}\}_{j,k=1}^{N_{\Phi },N_{\Lambda }}$ forms the set of $C_{\Phi }\varepsilon
^{1/(q+1)}$-brackets for $\mathcal{G}$ with respect to $||\cdot ||_{q}$.
This gives the desired entropy bound for $\mathcal{G}$. $\blacksquare $%
\bigskip 

We are prepared to present the uniform Bahadur representation of $\nu
_{n}(\lambda ,\varphi ,\psi )$. Let $\Lambda _{n}\equiv \{\lambda \in
\Lambda :||\lambda -\lambda _{0}||_{\infty }\leq c_{n}\}$, where $%
0<c_{n}n^{1/4}\rightarrow 0$. We let $X=[X_{1}^{\top },X_{2}^{\top }]^{\top
} $, where $X_{1}$ is a continuous random vector and $X_{2}$ is a discrete
random vector taking values in a finite set $\{x_{1},\cdot \cdot \cdot
,x_{M}\}$. We make the following assumptions.\bigskip

\noindent \textsc{Assumption B1} \textsc{:}\textbf{\ }(i) For some $C>0,\
p\geq q>4,$ $b_{\Psi }\in (0,q/(q-1)),\ $and $b_{\Phi }\in
(0,q/\{(q+1)(q-1)\}),$%
\begin{equation*}
\log N_{[]}(\varepsilon ,\Phi ,||\cdot ||_{p})<C\varepsilon ^{-b_{\Phi }}\ 
\text{and }\log N_{[]}(\varepsilon ,\Psi ,||\cdot ||_{p})<C\varepsilon
^{-b_{\Psi }},\ \text{for\ each\ }\varepsilon >0,
\end{equation*}%
and $\mathbf{E}[\tilde{\varphi}(W)^{p}]+\mathbf{E}[\tilde{\psi}%
(S)^{p}]+\sup_{x\in \mathcal{S}_{X}}\mathbf{E}[\tilde{\varphi}%
(W)|X=x]+\sup_{x\in \mathcal{S}_{X}}\mathbf{E}[\tilde{\psi}(W)|X=x]<\infty $.

\noindent (ii) (a) For $q>4$ in (i) and for some $b_{\Lambda }\in
(0,q/\{(q+1)(q-1)\})\ $and $C>0,$ 
\begin{equation*}
\log N_{[]}(\varepsilon ,\Lambda ,||\cdot ||_{\infty })\leq C\varepsilon
^{-b_{\Lambda }},
\end{equation*}%
$\ $for\ each\ $\varepsilon >0.$

\noindent (b) For all $\lambda \in \Lambda ,$ the density $f_{\lambda
}(\cdot )$ of $\lambda (X)$ is bounded uniformly over $\lambda \in \Lambda $
and bounded away from zero on the interior of its support uniformly over $%
\lambda \in \Lambda $.\bigskip

\noindent \textsc{Assumption B2} \textsc{:}\textbf{\ }(i) $K(\cdot )$ is
symmetric, nonnegative, compact supported, twice continuously differentiable
with bounded derivatives,\ and $\int K(t)dt=1$.

\noindent (ii) $n^{1/2}h^{3}+n^{-1/2}h^{-2}(-\log h)\rightarrow 0$ as $%
n\rightarrow \infty $.\bigskip

\noindent \textsc{Assumption B3} \textsc{:} $\mathbf{E}[\varphi
(W)|U_{\lambda }=\cdot ]$ is twice continuously differentiable with
derivatives bounded uniformly over $(\lambda ,\varphi )\in B(\lambda
_{0};\varepsilon )\times \Phi $ with some $\varepsilon >0.$\bigskip

The following lemma offers a uniform representation of $\nu _{n}.$\bigskip

\noindent \textsc{Lemma B2 :}\textbf{\ }\textit{Suppose that }Assumptions
B1-B3\textit{\ hold. Then,}%
\begin{equation*}
\sup_{(\lambda ,\varphi ,\psi )\in \Lambda _{n}\times \Phi \times \Psi
}\left\vert \nu _{n}(\lambda ,\varphi ,\psi )-\frac{1}{\sqrt{n}}%
\sum_{i=1}^{n}g_{\psi ,\lambda }(U_{\lambda ,i})\{\varphi (W_{i})-g_{\varphi
,\lambda }(U_{\lambda ,i})\}\right\vert =o_{P}(1)\text{\textit{.}}
\end{equation*}

\noindent \textsc{Proof of Lemma B2}: To make the flow of the arguments more
visible, the proof proceeds by making certain claims and proving them at the
end of the proof. Without loss of generality, assume that the support of $K$
is contained in $[-1,1].$ Throughout the proofs, the notation $\mathbf{E}_{i}
$ indicates conditional expectation given $(W_{i},S_{i},X_{i}).$

Let $\hat{\rho}_{\varphi ,\lambda ,i}(t)\equiv (n-1)^{-1}\sum_{j=1,j\neq
i}^{n}K_{h}(U_{n,\lambda ,j}-t)\varphi (W_{j}),$ $\xi _{1n}(u)\equiv
\int_{\lbrack -u/h,(1-u)/h]\cap \lbrack -1,1]}K(v)dv,$ and%
\begin{equation*}
\Delta _{i}^{\varphi ,\psi }(\lambda )\equiv g_{\psi ,\lambda }(U_{\lambda
,i})\{\varphi (W_{i})-g_{\varphi ,\lambda }(U_{\lambda ,i})\}.
\end{equation*}%
We write $\hat{g}_{\varphi ,\lambda ,i}(U_{n,\lambda ,i})-g_{\varphi
,\lambda }(U_{\lambda ,i})$ as%
\begin{eqnarray*}
R_{1i}(\lambda ,\varphi ) &\equiv &\frac{\hat{\rho}_{\varphi ,\lambda
,i}(U_{n,\lambda ,i})-g_{\varphi ,\lambda }(U_{\lambda ,i})\hat{f}_{\lambda
,i}(U_{\lambda ,i})}{\xi _{1n}(U_{\lambda ,i})} \\
&&+\frac{[\hat{\rho}_{\varphi ,\lambda ,i}(U_{n,\lambda ,i})-g_{\varphi
,\lambda }(U_{\lambda ,i})\hat{f}_{\lambda ,i}(U_{\lambda ,i})](\xi
_{1n}(U_{\lambda ,i})-\hat{f}_{\lambda ,i}(U_{n,\lambda ,i}))}{\hat{f}%
_{\lambda ,i}(U_{n,\lambda ,i})\xi _{1n}(U_{\lambda ,i})} \\
&=&R_{1i}^{A}(\lambda ,\varphi )+R_{1i}^{B}(\lambda ,\varphi ).
\end{eqnarray*}%
Put $\pi \equiv (\lambda ,\varphi ,\psi )$ and $\Pi _{n}\equiv \Lambda
_{n}\times \Phi \times \Psi ,$ and write%
\begin{equation*}
\nu _{n}(\pi )=\frac{1}{\sqrt{n}}\sum_{i=1}^{n}\psi
(S_{i})R_{1i}^{A}(\lambda ,\varphi )+\frac{1}{\sqrt{n}}\sum_{i=1}^{n}\psi
(S_{i})R_{1i}^{B}(\lambda ,\varphi )\equiv r_{1n}^{A}(\pi )+r_{1n}^{B}(\pi ),%
\text{ }\pi \in \Pi _{n}\text{.}
\end{equation*}%
First, we show the following:\bigskip

\noindent \textbf{C1:} $\sup_{\pi \in \Pi _{n}}\left\vert r_{1n}^{B}(\pi
)\right\vert =o_{P}(1).$\bigskip

We turn to $r_{1n}^{A}(\pi ),$ which we write as%
\begin{eqnarray*}
&&\frac{1}{(n-1)\sqrt{n}}\sum_{i=1}^{n}\sum_{j=1,j\neq i}^{n}\psi
_{n,\lambda ,i}\Delta _{\varphi ,\lambda ,ij}K_{ij}^{\lambda }+\frac{1}{(n-1)%
\sqrt{n}}\sum_{i=1}^{n}\sum_{j=1,j\neq i}^{n}\psi _{n,\lambda ,i}\Delta
_{\varphi ,\lambda ,ij}\{K_{n,ij}^{\lambda }-K_{ij}^{\lambda }\} \\
&\equiv &R_{1n}(\pi )+R_{2n}(\pi ),\text{ say,}
\end{eqnarray*}%
where $\psi _{n,\lambda ,i}\equiv \psi (S_{i})/\xi _{1n}(U_{\lambda ,i}),$ $%
\Delta _{\varphi ,\lambda ,ij}\equiv \varphi (W_{j})-g_{\varphi ,\lambda
}(U_{\lambda ,i}),$ $K_{n,ij}^{\lambda }\equiv K_{h}(U_{n,\lambda
,j}-U_{n,\lambda ,i})$ and $K_{ij}^{\lambda }\equiv K_{h}(U_{\lambda
,j}-U_{\lambda ,i}).$ We will now show that 
\begin{equation}
\text{sup}_{\pi \in \Pi _{n}}|R_{2n}(\pi )|\rightarrow _{P}0.  \label{stat}
\end{equation}

Let $\delta _{i}^{\lambda }\equiv U_{n,\lambda ,i}-U_{\lambda ,i}\ $and $%
d_{\lambda ,ji}\equiv \delta _{j}^{\lambda }-\delta _{i}^{\lambda }$ and
write $R_{2n}(\pi )$ as 
\begin{eqnarray*}
&&\frac{1}{(n-1)\sqrt{n}}\sum_{i=1}^{n}\sum_{j=1,j\neq i}^{n}\psi
_{n,\lambda ,i}\Delta _{\varphi ,\lambda ,ij}K_{h,ij}^{\prime }d_{\lambda
,ji}+\frac{1}{2(n-1)\sqrt{n}}\sum_{i=1}^{n}\sum_{j=1,j\neq i}^{n}\psi
_{n,\lambda ,i}\Delta _{\varphi ,\lambda ,ij}d_{\lambda
,ji}^{2}K_{h,ij}^{\prime \prime } \\
&=&A_{1n}(\pi )+A_{2n}(\pi ),\text{ say,}
\end{eqnarray*}%
where $K_{h,ij}^{\prime }\equiv h^{-2}\partial K(t)/\partial t$ at $%
t=(U_{\lambda ,i}-U_{\lambda ,j})/h\ $and 
\begin{equation*}
K_{h,ij}^{\prime \prime }\equiv h^{-3}\partial ^{2}K(t)/\partial
t^{2}|_{t=t_{ij}}
\end{equation*}%
with $t_{ij}\equiv \{(1-a_{ij})(U_{\lambda ,i}-U_{\lambda
,j})+a_{ij}(U_{n,\lambda ,i}-U_{n,\lambda ,j})\}/h,\ $for some $a_{ij}\in
\lbrack 0,1].\ $Later we will show the following:\bigskip

\noindent \textbf{C2:} sup$_{\pi \in \Pi _{n}}|A_{2n}(\pi )|=o_{P}(1).$%
\bigskip

We turn to $A_{1n}(\pi )\ $which we write as%
\begin{eqnarray}
&&\frac{1}{(n-1)\sqrt{n}}\sum_{i=1}^{n}\sum_{j=1,j\neq i}^{n}\psi
_{n,\lambda ,i}\Delta _{\varphi ,\lambda ,ij}K_{h,ij}^{\prime }\delta
_{j}^{\lambda }  \label{decomp5} \\
&&-\frac{1}{(n-1)\sqrt{n}}\sum_{i=1}^{n}\sum_{j=1,j\neq i}^{n}\psi
_{n,\lambda ,i}\Delta _{\varphi ,\lambda ,ij}K_{h,ij}^{\prime }\delta
_{i}^{\lambda }  \notag \\
&=&B_{1n}(\pi )+B_{2n}(\pi ),\text{ say.}  \notag
\end{eqnarray}%
Write $B_{1n}(\pi )$ as%
\begin{eqnarray*}
&&\frac{1}{n-1}\sum_{j=1}^{n}\left[ \frac{1}{\sqrt{n}}\sum_{i=1,i\neq
j}^{n}\left\{ \psi _{n,\lambda ,i}\Delta _{\varphi ,\lambda
,ij}K_{h,ij}^{\prime }-\mathbf{E}_{j}\left[ \psi _{n,\lambda ,i}\Delta
_{\varphi ,\lambda ,ij}K_{h,ij}^{\prime }\right] \right\} \right]
(U_{n,\lambda ,j}-U_{\lambda ,j}) \\
&&+\frac{1}{(n-1)\sqrt{n}}\sum_{i=1}^{n}\sum_{j=1,j\neq i}^{n}\mathbf{E}_{j}%
\left[ \psi _{n,\lambda ,i}\Delta _{\varphi ,\lambda ,ij}K_{h,ij}^{\prime }%
\right] (U_{n,\lambda ,j}-U_{\lambda ,j})=C_{1n}(\pi )+C_{2n}(\pi ),\text{
say.}
\end{eqnarray*}%
As for $C_{1n}(\pi ),$ we show the following later.\bigskip

\noindent \textbf{C3:} $\sup_{\pi \in \Pi _{n}}|C_{1n}(\pi )|=o_{P}(1).$%
\bigskip

We deduce a similar result for $B_{2n}(\pi )$, so that we write%
\begin{eqnarray}
A_{1n}(\pi ) &=&\frac{1}{(n-1)\sqrt{n}}\sum_{i=1}^{n}\sum_{j=1,j\neq i}^{n}%
\mathbf{E}_{j}\left[ \psi _{n,\lambda ,i}\Delta _{\varphi ,\lambda
,ij}K_{h,ij}^{\prime }\right] (U_{n,\lambda ,j}-U_{\lambda ,j})
\label{decomp6} \\
&&-\frac{1}{(n-1)\sqrt{n}}\sum_{j=1}^{n}\sum_{i=1,i\neq j}^{n}\mathbf{E}_{i}%
\left[ \psi _{n,\lambda ,i}\Delta _{\varphi ,\lambda ,ij}K_{h,ij}^{\prime }%
\right] (U_{n,\lambda ,i}-U_{\lambda ,i})+o_{P}(1)  \notag \\
&=&D_{1n}(\pi )-D_{2n}(\pi )+o_{P}(1)\text{, say.}  \notag
\end{eqnarray}%
Now, we show that $D_{1n}(\pi )$ and $D_{2n}(\pi )$ cancel out
asymptotically. As for $D_{1n}(\pi )$, using Hoeffding's decomposition and
taking care of the degenerate $U$-process (e.g. see C3 and its proof below),%
\begin{equation*}
\frac{1}{\sqrt{n}}\sum_{i=1}^{n}\int_{0}^{1}\mathbf{E}\left[ \psi
_{n,\lambda ,i}\Delta _{\varphi ,\lambda ,ij}K_{h,ij}^{\prime }\right]
\left( 1\{U_{\lambda ,i}\leq u_{1}\}-u_{1}\right) du_{1}+o_{P}(1),
\end{equation*}%
uniformly over $\pi \in \Pi _{n}$. Similarly, as for $D_{2n}(\pi )$, we can
write it as%
\begin{equation*}
\frac{1}{\sqrt{n}}\sum_{j=1}^{n}\int_{0}^{1}\mathbf{E}\left[ \psi
_{n,\lambda ,i}\Delta _{\varphi ,\lambda ,ij}K_{h,ij}^{\prime }\right]
\left( 1\{U_{\lambda ,j}\leq u_{1}\}-u_{1}\right) du_{1}+o_{P}(1),
\end{equation*}%
uniformly over $\pi \in \Pi _{n}$. Note that $\mathbf{E}\left[ \psi
_{n,\lambda ,i}\Delta _{\varphi ,\lambda ,ij}K_{h,ij}^{\prime }\right] $
does not depend on a particular choice of the pair $(i,j)$ as long as $i\neq
j$, so that%
\begin{equation*}
\mathbf{E}\left[ \psi _{n,\lambda ,i}\Delta _{\varphi ,\lambda
,ij}K_{h,ij}^{\prime }\right] =\mathbf{E}\left[ \psi _{n,\lambda ,j}\Delta
_{\varphi ,\lambda ,ji}K_{h,ji}^{\prime }\right] ,
\end{equation*}%
whenever $j\neq i$. Hence $D_{1n}(\pi )=D_{2n}(\pi )+o_{P}(1)$, uniformly
over $\pi \in \Pi _{n},\ $and that $\sup_{\pi \in \Pi _{n}}|A_{1n}(\pi
)|=o_{P}(1),$ which, together with (C2), completes the proof of (\ref{stat}).

It remains to show that%
\begin{equation}
\sup_{\pi \in \Pi _{n}}\left\vert R_{1n}(\pi )-\frac{1}{\sqrt{n}}%
\sum_{i=1}^{n}g_{\psi ,\lambda }(U_{\lambda ,i})\{\varphi (W_{i})-g_{\varphi
,\lambda }(U_{\lambda ,i})\}\right\vert =o_{P}(1).  \label{statement}
\end{equation}%
We define $q_{n,ij}^{\pi }\equiv \psi _{n,\lambda ,i}\Delta _{\varphi
,\lambda ,ij}K_{ij}^{\lambda }$ and write $R_{1n}(\pi )$ as%
\begin{equation}
\frac{1}{(n-1)\sqrt{n}}\sum_{i=1}^{n}\sum_{j=1,j\neq i}^{n}q_{n,ij}^{\pi }.
\label{ds}
\end{equation}%
Let $\rho _{n,ij}^{\pi }\equiv q_{n,ij}^{\pi }-\mathbf{E}_{i}[q_{n,ij}^{\pi
}]-\mathbf{E}_{j}[q_{n,ij}^{\pi }]+\mathbf{E}[q_{n,ij}^{\pi }]\ $and define%
\begin{equation*}
u_{n}(\pi )\equiv \frac{1}{(n-1)\sqrt{n}}\sum_{i=1}^{n}\sum_{j=1,j\neq
i}^{n}\rho _{n,ij}^{\pi }.
\end{equation*}%
Then, $\{u_{n}(\pi ):\pi \in \Pi _{n}\}$ is a degenerate $U$-process on $\Pi
_{n}$. We write (\ref{ds}) as%
\begin{equation}
\frac{1}{(n-1)\sqrt{n}}\sum_{i=1}^{n}\sum_{j=1,j\neq i}^{n}\left\{ \mathbf{E}%
_{i}[q_{n,ij}^{\pi }]+\mathbf{E}_{j}[q_{n,ij}^{\pi }]-\mathbf{E}%
[q_{n,ij}^{\pi }]\right\} +u_{n}(\pi ).  \label{decomp3}
\end{equation}%
We will later show the following two claims.\bigskip

\noindent \textbf{C4:} $\sup_{\pi \in \Pi _{n}}|\frac{1}{\sqrt{n}}%
\sum_{i=1}^{n}\{\mathbf{E}_{i}[q_{n,ij}^{\pi }]-\mathbf{E}[q_{n,ij}^{\pi
}]\}|=o_{P}(1).$

\noindent \textbf{C5: }$\sup_{\pi \in \Pi _{n}}|u_{n}(\pi )|=o_{P}(1).$%
\bigskip

We conclude from these claims that uniformly over $\pi \in \Pi _{n}$,%
\begin{equation*}
\frac{1}{(n-1)\sqrt{n}}\sum_{i=1}^{n}\sum_{j=1,j\neq i}^{n}q_{n,ij}^{\pi }=%
\frac{1}{\sqrt{n}}\sum_{j=1}^{n}\mathbf{E}_{j}[q_{n,ij}^{\pi }]+o_{P}(1).
\end{equation*}%
Then the proof of Lemma B2 is completed by showing the following.\bigskip

\noindent \textbf{C6:} $\sup_{\pi \in \Pi _{n}}\left\vert \frac{1}{\sqrt{n}}%
\sum_{j=1}^{n}\left( \mathbf{E}_{j}[q_{n,ij}^{\pi }]-g_{\psi ,\lambda
}(U_{\lambda ,j})\{\varphi (W_{j})-g_{\varphi ,\lambda }(U_{\lambda
,j})\}\right) \right\vert =o_{P}(1).$\bigskip

\noindent \textbf{Proof of C1:} From the proof of Lemma A3 of Song (2009)
(by replacing $\lambda $ with $F_{\lambda }\circ \lambda $ there), it
follows that%
\begin{equation}
\text{max}_{1\leq i\leq n}\text{sup}_{\lambda \in \Lambda }\text{sup}_{x\in 
\mathbf{R}^{d_{X}}}|F_{n,\lambda ,i}(\lambda (x))-F_{\lambda }(\lambda
(x))|=O_{P}(n^{-1/2}),  \label{rate}
\end{equation}%
where $F_{n,\lambda ,i}(\bar{\lambda})\equiv \frac{1}{n-1}\sum_{j=1,j\neq
i}^{n}1\{\lambda (X_{j})\leq \bar{\lambda}\}.$ We bound $\max_{1\leq i\leq
n}\sup_{\lambda \in \Lambda }|\hat{f}_{\lambda ,i}(U_{n,\lambda ,i})-\xi
_{1n}(U_{\lambda ,i})|$ by%
\begin{eqnarray}
&&\max_{1\leq i\leq n}\sup_{\lambda \in \Lambda }\sup_{u\in \lbrack
0,1]}\left\vert \hat{f}_{\lambda ,i}(u)-\xi _{1n}(u)\right\vert +\max_{1\leq
i\leq n}\sup_{\lambda \in \Lambda }\left\vert \xi _{1n}(U_{n,\lambda
,i})-\xi _{1n}(U_{\lambda ,i})\right\vert   \label{conv} \\
&=&O_{P}(n^{-1/2}h^{-1}\sqrt{-\log h}%
)+O_{P}(n^{-1/2}h^{-1})=O_{P}(n^{-1/2}h^{-1}\sqrt{-\log h}),  \notag
\end{eqnarray}%
using (29) of Song (2009). Hence, uniformly over $1\leq i\leq n$ and over $%
\lambda \in \Lambda ,$%
\begin{eqnarray*}
&&|\hat{\rho}_{\varphi ,\lambda ,i}(U_{n,\lambda ,i})-\hat{f}_{\lambda
,i}(U_{n,\lambda ,i})g_{\varphi ,\lambda }(U_{\lambda ,i})| \\
&\leq &|\hat{\rho}_{\varphi ,\lambda ,i}(U_{n,\lambda ,i})-\xi
_{1n}(U_{\lambda ,i})g_{\varphi ,\lambda }(U_{\lambda ,i})|+|g_{\varphi
,\lambda }(U_{\lambda ,i})||\xi _{1n}(U_{\lambda ,i})-\hat{f}_{\lambda
,i}(U_{n,\lambda ,i})| \\
&=&|\hat{\rho}_{\varphi ,\lambda ,i}(U_{n,\lambda ,i})-\xi _{1n}(U_{\lambda
,i})g_{\varphi ,\lambda }(U_{\lambda ,i})|+O_{P}(n^{-1/2}h^{-1}\sqrt{-\log h}%
).
\end{eqnarray*}%
As for the leading term, we apply (23) of Song (2009) and Assumption B3 to
deduce that uniformly over $1\leq i\leq n$ and over $(\lambda ,\varphi )\in
\Lambda \times \Phi $,%
\begin{eqnarray}
|\hat{\rho}_{\varphi ,\lambda ,i}(U_{n,\lambda ,i})-\xi _{1n}(U_{\lambda
,i})g_{\varphi ,\lambda }(U_{\lambda ,i})|(1-1_{n,\lambda ,i}) &=&O_{P}(h)%
\text{ and}  \label{conv2} \\
|\hat{\rho}_{\varphi ,\lambda ,i}(U_{n,\lambda ,i})-\xi _{1n}(U_{\lambda
,i})g_{\varphi ,\lambda }(U_{\lambda ,i})|1_{n,\lambda ,i}
&=&O_{P}(h^{2}+n^{-1/2}h^{-1}\sqrt{-\log h}),  \notag
\end{eqnarray}%
where $1_{n,\lambda ,i}=1\left\{ |1-U_{n,\lambda ,i}|>h\right\} $. Also,
observe that (e.g. see arguments after (29) of Song (2009)) 
\begin{equation}
\xi _{1n}(u)\geq 1/2\ \text{for\ all\ }u\in \lbrack 0,1].  \label{lb}
\end{equation}%
Therefore, we bound $\left\vert r_{1n}^{B}(\pi )\right\vert $ by 
\begin{eqnarray*}
&&\frac{C_{1}}{\sqrt{n}}\sum_{i=1}^{n}\left\vert [\hat{\rho}_{\varphi
,\lambda ,i}(U_{n,\lambda ,i})-g_{\varphi ,\lambda }(U_{\lambda ,i})\xi
_{1n}(U_{\lambda ,i})](\xi _{1n}(U_{\lambda ,i})-\hat{f}_{\lambda
,i}(U_{n,\lambda ,i}))\right\vert 1_{n,\lambda ,i} \\
&&+\frac{C_{2}}{\sqrt{n}}\sum_{i=1}^{n}\left\vert [\hat{\rho}_{\varphi
,\lambda ,i}(U_{n,\lambda ,i})-g_{\varphi ,\lambda }(U_{\lambda ,i})\xi
_{1n}(U_{\lambda ,i})](\xi _{1n}(U_{\lambda ,i})-\hat{f}_{\lambda
,i}(U_{n,\lambda ,i}))\right\vert (1-1_{n,\lambda ,i}),
\end{eqnarray*}%
for some $C_{1},C_{2}>0$, uniformly over $\pi \in \Pi _{n}$ with large
probability. From (\ref{conv}) and (\ref{conv2}), the first term is equal to 
$O_{P}(n^{-1/2}h^{-2}(-\log h))=o_{P}(1)$ uniformly over$\ 1\leq i\leq n$.
As for the second term, from (\ref{conv}) and (\ref{conv2}), it is bounded
with large probability by%
\begin{equation*}
Cn^{-1/2}\sqrt{-\log h}\cdot \frac{1}{n}\sum_{i=1}^{n}1\left\{
|1-U_{0,i}|\leq Ch\right\} =O_{P}(n^{-1/2}h\sqrt{-\log h})=o_{P}(n^{-1/2}),
\end{equation*}%
for some $C>0$ with large probability. Hence (C1) is established.\bigskip 

\noindent \textbf{Proof of C2:} Let $\tilde{\Delta}_{ij}\equiv \tilde{\varphi%
}(W_{i})+\sup_{x\in \mathcal{S}_{X}}\mathbf{E}[\tilde{\varphi}%
(W_{j})|X_{j}=x].\ $Since max$_{1\leq i,j\leq n}\sup_{\lambda \in \Lambda
}d_{\lambda ,ji}^{2}=O_{P}(n^{-1})\ $by (\ref{rate}), with large
probability, we bound $|A_{2n}(\pi )|$ by the absolute value of%
\begin{equation}
\frac{C}{(n-1)n\sqrt{n}}\sum_{i=1}^{n}\sum_{j=1,j\neq i}^{n}\left( \tilde{%
\psi}_{i}\tilde{\Delta}_{ij}\left\vert K_{h,ij}^{\prime \prime }\right\vert -%
\mathbf{E}\left[ \tilde{\psi}_{i}\tilde{\Delta}_{ij}\left\vert
K_{h,ij}^{\prime \prime }\right\vert \right] \right) +\frac{\sqrt{n}}{n}%
\mathbf{E}\left[ \tilde{\psi}_{i}\tilde{\Delta}_{ij}\left\vert
K_{h,ij}^{\prime \prime }\right\vert \right] .  \label{dec5}
\end{equation}%
Using the standard \textit{U} statistics theory, we bound the leading term by%
\begin{eqnarray*}
&&\left\vert \frac{C}{n\sqrt{n}}\sum_{i=1}^{n}\left( \mathbf{E}_{i}\left[ 
\tilde{\psi}_{i}\tilde{\Delta}_{ij}\left\vert K_{h,ij}^{\prime \prime
}\right\vert \right] -\mathbf{E}\left[ \tilde{\psi}_{i}\tilde{\Delta}%
_{ij}\left\vert K_{h,ij}^{\prime \prime }\right\vert \right] \right)
\right\vert  \\
&&+\left\vert \frac{C}{n\sqrt{n}}\sum_{j=1}^{n}\left( \mathbf{E}_{j}\left[ 
\tilde{\psi}_{i}\tilde{\Delta}_{ij}\left\vert K_{h,ij}^{\prime \prime
}\right\vert \right] -\mathbf{E}\left[ \tilde{\psi}_{i}\tilde{\Delta}%
_{ij}\left\vert K_{h,ij}^{\prime \prime }\right\vert \right] \right)
\right\vert +o_{P}(1).
\end{eqnarray*}%
The expected value of the sum above is bounded by%
\begin{equation*}
\frac{C}{n}\sqrt{Var\left( \mathbf{E}_{i}\left[ \tilde{\psi}_{i}\tilde{\Delta%
}_{ij}\left\vert K_{h,ij}^{\prime \prime }\right\vert \right] \right) }+%
\frac{C}{n}\sqrt{Var\left( \mathbf{E}_{j}\left[ \tilde{\psi}_{i}\tilde{\Delta%
}_{ij}\left\vert K_{h,ij}^{\prime \prime }\right\vert \right] \right) }%
=O(n^{-1}h^{-5/2})=o(1),
\end{equation*}%
so that the leading term of (\ref{dec5}) is $o(1)$. On the other hand, $%
\frac{\sqrt{n}}{n}\mathbf{E}\left[ \tilde{\psi}_{i}\tilde{\Delta}%
_{ij}|K_{h,ij}^{\prime \prime }|\right] =O(n^{-1/2}h^{-2})=o(1).$\bigskip 

\noindent \textbf{Proof of C3:} Let $K^{\prime }(u)\equiv (\partial
K(u)/\partial u)1\{u\in (0,1)\}.$ Then $K^{\prime }(\cdot /h)$ is uniformly
bounded and bounded variation. Let $\mathcal{K}_{\Lambda }\equiv \{K^{\prime
}((u-\sigma (\cdot ))/h):(\sigma ,u)\in \mathcal{I}\times \lbrack 0,1]\},$
where $\mathcal{I}\equiv \{(F_{\lambda }\circ \lambda )(x):\lambda \in
\Lambda \}$. Also, let $\mathcal{W}_{\Lambda }\equiv \{\xi _{1n}(\sigma
(\cdot )):\sigma \in \mathcal{I}\}$. Take $p\geq 4$ as in Assumption B1. By
Lemma A1 of Song (2009) and Assumptions B1(ii) and B3(iv),%
\begin{eqnarray}
\log N_{[]}(\varepsilon ,\mathcal{K}_{\Lambda },||\cdot ||_{p}) &\leq &\log
N(C\varepsilon ,\mathcal{I},||\cdot ||_{\infty })+C/\varepsilon \leq
C\varepsilon ^{-b_{\Lambda }}\text{ and}  \label{beb} \\
\log N_{[]}(\varepsilon ,\mathcal{W}_{\Lambda },||\cdot ||_{p}) &\leq &\log
N(C\varepsilon ,\mathcal{I},||\cdot ||_{\infty })+C/\varepsilon \leq
C\varepsilon ^{-b_{\Lambda }},  \notag
\end{eqnarray}%
for some $C>0$. Let $\xi _{\pi ,u}(S_{i},X_{i})\equiv \psi _{n,\lambda
,i}K^{\prime }((u-U_{\lambda ,i})/h)$. We bound $\sup_{\pi \in \Pi
_{n}}|C_{1n}(\pi )|\leq h^{-2}G_{1n}\cdot V_{1n}+h^{-2}G_{2n}\cdot V_{2n}$,
where%
\begin{eqnarray*}
V_{1n} &\equiv &\sup_{(\pi ,u)\in \Pi _{n}\times \lbrack 0,1]}\left\vert 
\frac{1}{\sqrt{n}}\sum_{i=1}^{n}\left( \xi _{\pi ,u}(S_{i},X_{i})-\mathbf{E}%
\xi _{\pi ,u}(S_{i},X_{i})\right) \right\vert \text{ and} \\
V_{2n} &\equiv &\sup_{(\pi ,u)\in \Pi _{n}\times \lbrack 0,1]}\left\vert 
\frac{1}{\sqrt{n}}\sum_{i=1}^{n}\left( \xi _{\pi ,u}(S_{i},X_{i})g_{\varphi
,\lambda }(U_{\lambda ,i})-\mathbf{E}\xi _{\pi ,u}(S_{i},X_{i})g_{\varphi
,\lambda }(U_{\lambda ,i})\right) \right\vert ,
\end{eqnarray*}%
where $G_{1n}\equiv \sup_{\lambda \in \Lambda _{n}}\frac{1}{n}\sum_{j=1}^{n}%
\tilde{\varphi}(W_{j})|\delta _{j}^{\lambda }|$ and $G_{2n}\equiv
\sup_{\lambda \in \Lambda _{n}}\frac{1}{n}\sum_{j=1}^{n}|\delta
_{j}^{\lambda }|$. Define 
\begin{eqnarray*}
\mathcal{F}_{1} &\equiv &\{\xi _{\pi ,u}(\cdot ,\cdot ):(\pi ,u)\in \Pi
_{n}\times \lbrack 0,1]\}\text{ and} \\
\mathcal{F}_{2} &\equiv &\{\xi _{\pi ,u}(\cdot ,\cdot )\cdot (g_{\varphi
,\lambda }\circ \sigma _{\lambda })(\cdot ):(\pi ,u)\in \Pi _{n}\times
\lbrack 0,1]\}.
\end{eqnarray*}%
Then, $\mathcal{F}_{1}\subset (\Psi /\mathcal{W}_{\Lambda })\cdot \mathcal{K}%
_{1,\Lambda }$ and $\mathcal{F}_{2}\subset (\Psi /\mathcal{W}_{\Lambda
})\cdot \mathcal{K}_{1,\Lambda }\cdot \mathcal{H}$, where%
\begin{equation*}
\mathcal{H}\equiv \{(g_{\varphi ,\lambda }\circ \sigma _{\lambda })(\cdot
):(\varphi ,\lambda )\in \Phi \times \Lambda \}.
\end{equation*}%
By Assumption B1(i) and Lemma B1, $\log N_{[]}(\varepsilon ,\mathcal{H}%
,||\cdot ||_{q})\leq C\varepsilon ^{-(q+1)\{b_{\Phi }\vee b_{\Lambda }\}}.$
Combining this with (\ref{beb}), the entropy bound for $\Psi $ in Assumption
B1(i), (\ref{beb}), and the fact that $\sup_{x\in \mathcal{S}_{X}}\mathbf{E}%
\left[ \tilde{\psi}(S)|X=x\right] <\infty $ and $\sup_{x\in \mathcal{S}_{X}}%
\mathbf{E}\left[ \tilde{\varphi}(W)|X=x\right] <\infty $, we find that%
\begin{equation*}
\log N_{[]}(\varepsilon ,\mathcal{F}_{1},||\cdot ||_{2})\leq C\varepsilon
^{-\{b_{\Psi }\vee b_{\Phi }\vee b_{\Lambda }\}}\text{, and }\log
N_{[]}(\varepsilon ,\mathcal{F}_{2},||\cdot ||_{2})\leq C\varepsilon
^{-(b_{\Psi }\vee \lbrack (q+1)\{b_{\Phi }\vee b_{\Lambda }\}])}.
\end{equation*}%
From (\ref{lb}), we take an envelope of $\mathcal{F}_{1}$ as $\bar{\xi}%
_{1}(s,x)=2\tilde{\psi}(s)||K^{\prime }||_{\infty }$ and an envelope of $%
\mathcal{F}_{2}$ as $\bar{\xi}_{2}(s,x)=2\tilde{\psi}(s)||K^{\prime
}||_{\infty }\cdot \sup_{x\in \mathcal{S}_{X}}\mathbf{E}\left[ \tilde{\varphi%
}(W)|X=x\right] $. Certainly, $||\bar{\xi}_{1}||_{p}<\infty $ and $||\bar{\xi%
}_{2}||_{p}<\infty $ by Assumption B1(i). Using the maximal inequality of
Pollard (1989)\footnote{%
The result is replicated in Theorem A.2 of van der Vaart (1996).} and using
the fact that $b_{\Psi }\vee \lbrack (q+1)\{b_{\Phi }\vee b_{\Lambda }\}]<2$
(Assumptions B1(i) and B1(ii)(a)), we find that%
\begin{equation*}
V_{1n}=O_{P}(1)\text{ and\ }V_{2n}=O_{P}(1).
\end{equation*}%
By the fact that $\max_{1\leq j\leq n}\sup_{\lambda \in \Lambda _{n}}|\delta
_{j}^{\lambda }|=O_{P}(n^{-1/2}),$ we also deduce that $%
G_{1n}=O_{P}(n^{-1/2})$ and $G_{2n}=O_{P}(n^{-1/2})$. The desired result
follows because $O_{P}(n^{-1/2}h^{-2})=o_{P}(1).$\bigskip 

\noindent \textbf{Proof of C4:}\ Observe that $\mathbf{E}_{i}[q_{n,ij}^{\pi
}]$ is equal to $\zeta _{\pi }(S_{i},X_{i})$, where for $\pi \in \Pi _{n},$%
\begin{equation*}
\zeta _{\pi }(S_{i},X_{i})=\frac{\psi (S_{i})}{\xi _{1n}(U_{\lambda ,i})}%
\int_{0}^{1}\{g_{\varphi ,\lambda }(u)-g_{\varphi ,\lambda }(U_{\lambda
,i})\}K_{h}(u-U_{\lambda ,i})du.
\end{equation*}%
Define $\mathcal{F}_{3}=\{\zeta _{\pi }(\cdot ,\cdot ):\pi \in \Pi _{n}\}$.
Take $p\geq 4$ as in Assumption B1. Then, similarly as in the proof of (C3),
we can show that%
\begin{equation}
\log N_{[]}(\varepsilon ,\mathcal{F}_{3},||\cdot ||_{q})\leq C\varepsilon
^{-(b_{\Psi }\vee \lbrack (q+1)\{b_{\Phi }\vee b_{\Lambda }\}])}  \label{ent}
\end{equation}%
for some $C>0$. With $\sigma _{\lambda }(x)\equiv (F_{\lambda }\circ \lambda
)(x)$, we take%
\begin{equation*}
\bar{\zeta}(s,x)=2\tilde{\psi}(s)\sup_{(\varphi ,\lambda )\in \Phi \times
\Lambda }\left\vert \int_{0}^{1}\{g_{\varphi ,\lambda }(u)-g_{\varphi
,\lambda }(\sigma _{\lambda }(x))\}K_{h}(u-\sigma _{\lambda
}(x))du\right\vert 
\end{equation*}%
as an envelope of $\mathcal{F}_{3}$. Observe that $\mathbf{E}[\bar{\zeta}%
(S_{i},X_{i})^{2}]$ is bounded by%
\begin{equation}
4\sup_{x\in \mathcal{S}_{X}}\mathbf{E}\left[ \tilde{\psi}^{2}(S)|X=x\right]
\cdot \int_{0}^{1}\sup_{(\varphi ,\lambda )\in \Phi \times \Lambda }\left[
\int_{0}^{1}\{g_{\varphi ,\lambda }(t_{2})-g_{\varphi ,\lambda
}(t_{1})\}K_{h}(t_{2}-t_{1})dt_{2}\right] ^{2}dt_{1}.  \label{ex2}
\end{equation}%
By change of variables, the integral inside the bracket becomes%
\begin{equation*}
\int_{\lbrack -t_{1}/h,(1-t_{1})/h]\cap \lbrack -1,1]}\{g_{\varphi ,\lambda
}(t_{1}+ht_{2})-g_{\varphi ,\lambda }(t_{1})\}K(t_{2})dt_{2}.
\end{equation*}%
After tedious algebra (using the symmetry of $K$), we can show that the
outside integral in (\ref{ex2})\ is $O(h^{3}).$ Therefore, $||\bar{\zeta}%
||_{2}=O(h^{3/2})$ as $n\rightarrow \infty .\ $Applying this and the maximal
inequality of Pollard (1989),%
\begin{equation*}
\mathbf{E}\left[ \sup_{\pi \in \Pi _{n}}\left\vert \frac{1}{\sqrt{n}}%
\sum_{i=1}^{n}(\mathbf{E}_{i}[q_{n,ij}^{\pi }]-\mathbf{E}[q_{n,ij}^{\pi
}])\right\vert \right] \leq C\int_{0}^{Ch^{3/2}}\sqrt{1+\log
N_{[]}(\varepsilon ,\mathcal{F}_{3},||\cdot ||_{2})}d\varepsilon 
\end{equation*}%
for some $C>0$. By (\ref{ent}), the last term is of order $O(h^{(3/2)\times
\{1-(b_{\Psi }\vee \lbrack (q+1)\{b_{\Phi }\vee b_{\Lambda }\}])/2\}}\sqrt{%
-\log h})$. Since $b_{\Psi }\vee \lbrack (q+1)\{b_{\Phi }\vee b_{\Lambda
}\}]<2,$ we obtain the wanted result.\bigskip 

\noindent \textbf{Proof of C5:\ }Let us define $\mathcal{\tilde{J}}%
_{n}=\{hq_{n}^{\pi }(\cdot ,\cdot ):\pi \in \Pi \},\ $where $q_{n}^{\pi
}(Z_{i},Z_{j})=q_{n,ij}^{\pi },$ $Z_{i}=(S_{i},W_{i},X_{i})$, and $%
q_{n,ij}^{\pi }$ is defined prior to (\ref{ds}). Take $q>4$ as in Assumption
B1. Using similar arguments as in (C3), we can show that for some $C>0$, $%
\log N_{[]}(\varepsilon ,\mathcal{\tilde{J}}_{n},||\cdot ||_{q})\leq
C\varepsilon ^{-(b_{\Psi }\vee \lbrack (q+1)\{b_{\Phi }\vee b_{\Lambda
}\}])} $ for all $\varepsilon >0$. By Assumption B1, $(b_{\Psi }\vee \lbrack
(q+1)\{b_{\Phi }\vee b_{\Lambda }\}])(1-1/q)<1$. Then, from the proof of C3,%
\begin{equation*}
\int_{0}^{1}\left\{ \log N_{[]}(\varepsilon ,\mathcal{\tilde{J}}_{n},||\cdot
||_{q})\right\} ^{(1-1/q)}d\varepsilon \leq C\int_{0}^{1}\varepsilon
^{-(b_{\Psi }\vee (q+1)\{b_{\Phi }\vee b_{\Lambda }\})(1-1/q)}d\varepsilon
<\infty
\end{equation*}%
for some $C>0$. Furthermore, as in the proof of C3, we can take an envelope
of $\mathcal{\tilde{J}}_{n}$ that is $L_{q}$-bounded. By Theorem 1 of
Turki-Moalla (1998), p.878, for some small $\varepsilon >0,$%
\begin{equation*}
h\sup_{\pi \in \Pi _{n}}|u_{1n}(\pi )|=O_{P}(n^{1/2-(1-1/q)+\varepsilon
})=O_{P}(n^{-1/2+1/q+\varepsilon }).
\end{equation*}%
Therefore,\ $\sup_{\pi \in \Pi _{n}}|u_{1n}(\pi
)|=O_{P}(n^{-1/2+1/q+\varepsilon }h^{-1})=o_{P}(1)$ by taking small $%
\varepsilon >0$ and using Assumption B2(ii) and the fact that $q>4$. Hence
the proof is complete.\bigskip

\noindent \textbf{Proof of C6}: For $i\neq j,$ we write%
\begin{eqnarray*}
&&\frac{1}{\sqrt{n}}\sum_{j=1}^{n}\left( \mathbf{E}_{j}[q_{n,ij}^{\pi
}]-g_{\psi ,\lambda }(U_{\lambda ,j})\{\varphi (W_{j})-g_{\varphi ,\lambda
}(U_{\lambda ,j})\}\right)  \\
&=&\frac{1}{\sqrt{n}}\sum_{j=1}^{n}\left( \mathbf{E}_{j}\left[ \left(
q_{n,ij}^{\pi }-\frac{1}{\xi _{1n}(U_{\lambda ,i})}g_{\psi ,\lambda
}(U_{\lambda ,j})\{\varphi (W_{j})-g_{\varphi ,\lambda }(U_{\lambda
,j})\}\right) \right] \right)  \\
&&-\mathbf{E}\left[ \left( \frac{\xi _{1n}(U_{\lambda ,i})-1}{\xi
_{1n}(U_{\lambda ,i})}\right) \right] \times \frac{1}{\sqrt{n}}%
\sum_{j=1}^{n}g_{\psi ,\lambda }(U_{\lambda ,j})\{\varphi (W_{j})-g_{\varphi
,\lambda }(U_{\lambda ,j})\} \\
&\equiv &E_{1n}(\pi )-E_{2n}(\pi )\text{, say.}
\end{eqnarray*}%
We focus on $E_{1n}$ first. Note that%
\begin{eqnarray}
&&\mathbf{E}\left[ \sup_{\pi \in \Pi _{n}}\left( \mathbf{E}%
_{j}[q_{n,ij}^{\pi }]-\frac{g_{\psi ,\lambda }(U_{\lambda ,j})\{\varphi
(W_{j})-g_{\varphi ,\lambda }(U_{\lambda ,j})\}}{\xi _{1n}(U_{\lambda ,i})}%
\right) ^{2}\right]   \label{term2} \\
&=&\int \sup_{\pi \in \Pi _{n}}\left\{ \int_{0}^{1}A_{n,\pi
}(t_{1},t_{2},y)dt_{1}\right\} ^{2}dF_{Y,\lambda }(y,t_{2}),  \notag
\end{eqnarray}%
where $\int \cdot dF_{Y,\lambda }$ denotes the integration with respect to
the joint distribution of $(W_{i},U_{\lambda ,i})$ and%
\begin{equation*}
A_{n,\pi }(t_{1},t_{2},y)=\frac{1}{\xi _{1n}(t_{1})}\left( g_{\psi ,\lambda
}(t_{1})\{\varphi (y)-g_{\varphi ,\lambda
}(t_{1})\}K_{h}(t_{1}-t_{2})-g_{\psi ,\lambda }(t_{2})\{\varphi
(y)-g_{\varphi ,\lambda }(t_{2})\}\right) .
\end{equation*}%
After some tedious algebra, we can show that the last term in (\ref{term2})
is $O(h^{3})$ (see the proof of C4). Using the similar arguments as in the
proof of (C4), we find that 
\begin{equation*}
\text{sup}_{\pi \in \Pi _{n}}|E_{1n}(\pi )|=o_{P}(1).
\end{equation*}

We turn to $E_{2n}$. It is not hard to see that for all $\lambda \in \Lambda
_{n},$%
\begin{equation*}
\mathbf{E}\left[ \left( \frac{\xi _{1n}(U_{\lambda ,i})-1}{\xi
_{1n}(U_{\lambda ,i})}\right) \right] =\mathbf{E}\left[ \left( \frac{\xi
_{1n}(U_{0,i})-1}{\xi _{1n}(U_{0,i})}\right) \right] =o(1),\text{ as }%
h\rightarrow 0.
\end{equation*}%
The first equality follows because $U_{\lambda ,i}$ follows a uniform
distribution on $[0,1]$ for all $\lambda \in \Lambda $. Furthermore, we have%
\begin{equation*}
\sup_{\pi \in \Pi _{n}}\left\vert \frac{1}{\sqrt{n}}\sum_{j=1}^{n}\left(
g_{\psi ,\lambda }(U_{\lambda ,j})\{\varphi (W_{j})-g_{\varphi ,\lambda
}(U_{\lambda ,j})\}\right) \right\vert =O_{P}(1),
\end{equation*}%
using bracketing entropy conditions for $\Psi ,$ $\Phi $, and $\Lambda _{n}$%
, Lemma B1, and the maximal inequality of Pollard (1989). Therefore, we find
that sup$_{\pi \in \Pi _{n}}|E_{2n}(\pi )|=o_{P}(1)$. $\blacksquare $%
\bigskip 

Let $D_{i}\in \{0,1\}$ be a binary random variable and for $d\in \{0,1\}$,
define $g_{\varphi ,\lambda ,d}(u)\equiv \mathbf{E}[\varphi
(W_{i})|U_{\lambda ,i}=u,D_{i}=d]$ and $g_{\psi ,\lambda ,d}(u)\equiv 
\mathbf{E}[\psi (S_{i})|U_{\lambda ,i}=u,D_{i}=d].$ Consider the estimator:%
\begin{equation*}
\hat{g}_{\varphi ,\lambda ,d}(U_{n,\lambda ,i})\equiv \frac{1}{(n-1)\hat{f}%
_{\lambda ,d}(U_{n,\lambda ,i})}\sum_{j=1,j\neq i}^{n}\varphi
(W_{j})1\{D_{j}=d\}K_{h}\left( U_{n,\lambda ,j}-U_{n,\lambda ,i}\right) ,
\end{equation*}%
where $\hat{f}_{\lambda ,d}(U_{n,\lambda ,i})\equiv
(n-1)^{-1}\sum_{j=1,j\neq i}^{n}1\{D_{j}=d\}K_{h}(U_{n,\lambda
,j}-U_{n,\lambda ,i}).$ Similarly as before, we define%
\begin{equation*}
\nu _{n,d}(\lambda ,\varphi ,\psi )\equiv \frac{\sqrt{n}}{\sum_{i=1}^{n}D_{i}%
}\sum_{i=1}^{n}\psi (S_{i})D_{i}\left\{ \hat{g}_{\varphi ,\lambda
,d}(U_{n,\lambda ,i})-g_{\varphi ,\lambda ,d}(U_{\lambda ,i})\right\} ,
\end{equation*}%
with $(\lambda ,\varphi ,\psi )\in \Lambda \times \Phi \times \Psi .$ The
following lemma presents variants of Lemma B2.\bigskip

\noindent \textsc{Lemma B3} \textbf{: }\textit{Suppose that }Assumptions
B1-B3\textit{\ hold, and let }$P_{d}\equiv P\{D=d\}$\textit{, and }$%
\varepsilon _{\varphi ,\lambda ,d,i}\equiv \varphi (W_{i})-g_{\varphi
,\lambda ,d}(U_{\lambda ,i})$, $d\in \{0,1\}.$

\noindent (i) \textit{If there exists }$\varepsilon >0$ \textit{such that }$%
P\{D_{i}=1|U_{\lambda ,i}=u\}\geq \varepsilon $ \textit{for all} $(u,\lambda
)\in \lbrack 0,1]\times \Lambda ,$ \textit{then}%
\begin{equation*}
\sup_{(\lambda ,\varphi ,\psi )\in \Lambda _{n}\times \Phi \times \Psi
}\left\vert \nu _{n,1}(\lambda ,\varphi ,\psi )-\frac{1}{\sqrt{n}P_{1}}%
\sum_{i=1}^{n}D_{i}g_{\psi ,\lambda ,1}(U_{\lambda ,i})\varepsilon _{\varphi
,\lambda ,1,i}\right\vert =o_{P}(1)\text{\textit{.}}
\end{equation*}

\noindent (ii) \textit{If there exists }$\varepsilon >0$ \textit{such that }$%
P\{D_{i}=1|U_{\lambda ,i}=u\}\in \lbrack \varepsilon ,1-\varepsilon ]$ 
\textit{for all} $(u,\lambda )\in \lbrack 0,1]\times \Lambda ,$ \textit{then}%
\begin{equation*}
\sup_{(\lambda ,\varphi ,\psi )\in \Lambda _{n}\times \Phi \times \Psi
}\left\vert \nu _{n,0}(\lambda ,\varphi ,\psi )-\frac{1}{\sqrt{n}P_{1}}%
\sum_{i=1}^{n}\frac{(1-D_{i})P(U_{\lambda ,i})g_{\psi ,\lambda
,1}(U_{\lambda ,i})}{1-P(U_{\lambda ,i})}\varepsilon _{\varphi ,\lambda
,0,i}\right\vert =o_{P}(1).
\end{equation*}

\noindent \textsc{Proof of Lemma B3:} (i) Write%
\begin{equation*}
\nu _{n,1}(\lambda ,\varphi ,\psi )=\frac{1}{\frac{1}{n}\sum_{i=1}^{n}D_{i}}%
\cdot \frac{1}{\sqrt{n}}\sum_{i=1}^{n}\psi (S_{i})D_{i}\left\{ \frac{\hat{g}%
_{\varphi ,\lambda ,i}^{[1]}(U_{n,\lambda ,i})}{\hat{g}_{\lambda
,i}^{[1]}(U_{n,\lambda ,i})}-\frac{g_{\varphi ,\lambda }^{[1]}(U_{\lambda
,i})}{g_{\lambda }^{[1]}(U_{\lambda ,i})}\right\} ,
\end{equation*}%
where $g_{\varphi ,\lambda }^{[d]}(u)=\mathbf{E}[\varphi
(W_{i})1\{D_{i}=d\}|U_{\lambda ,i}=u]$, $g_{\lambda
}^{[d]}(u)=P\{D_{i}=d|U_{\lambda ,i}=u\}$,%
\begin{eqnarray*}
\hat{g}_{\varphi ,\lambda ,i}^{[d]}(u) &=&\frac{1}{(n-1)\hat{f}_{\lambda
,d}(u)}\sum_{j=1,j\neq i}^{n}\varphi (W_{j})1\{D_{j}=d\}K_{h}\left(
U_{n,\lambda ,j}-u\right) \text{, and} \\
\hat{g}_{\lambda ,i}^{[d]}(u) &=&\frac{1}{(n-1)\hat{f}_{\lambda ,d}(u)}%
\sum_{j=1,j\neq i}^{n}1\{D_{j}=d\}K_{h}\left( U_{n,\lambda ,j}-u\right) .
\end{eqnarray*}%
Using the arguments in the proof of Lemma B2, we can write%
\begin{eqnarray*}
&&\frac{1}{\sqrt{n}}\sum_{i=1}^{n}\psi (S_{i})D_{i}\left\{ \frac{\hat{g}%
_{\varphi ,\lambda ,i}^{[1]}(U_{n,\lambda ,i})}{\hat{g}_{\lambda
,i}^{[1]}(U_{n,\lambda ,i})}-\frac{g_{\varphi ,\lambda }^{[1]}(U_{\lambda
,i})}{g_{\lambda }^{[1]}(U_{\lambda ,i})}\right\}  \\
&=&\frac{1}{\sqrt{n}}\sum_{i=1}^{n}\frac{\psi (S_{i})D_{i}}{g_{\lambda
}^{[1]}(U_{\lambda ,i})}\left\{ \hat{g}_{\varphi ,\lambda
,i}^{[1]}(U_{n,\lambda ,i})-g_{\varphi ,\lambda }^{[1]}(U_{\lambda
,i})\right\}  \\
&&+\frac{1}{\sqrt{n}}\sum_{i=1}^{n}\psi (S_{i})D_{i}\frac{g_{\varphi
,\lambda }^{[1]}(U_{\lambda ,i})}{(g_{\lambda }^{[1]}(U_{\lambda ,i}))^{2}}%
\left\{ g_{\lambda }^{[1]}(U_{\lambda ,i})-\hat{g}_{\lambda
,i}^{[1]}(U_{n,\lambda ,i})\right\} +o_{P}(1).
\end{eqnarray*}%
By applying Lemma B2 to both terms, we obtain that%
\begin{eqnarray*}
&&\frac{1}{\sqrt{n}}\sum_{i=1}^{n}\psi (S_{i})D_{i}\left\{ \frac{\hat{g}%
_{\varphi ,\lambda ,i}^{[1]}(U_{n,\lambda ,i})}{\hat{g}_{\lambda
,i}^{[1]}(U_{n,\lambda ,i})}-\frac{g_{\varphi ,\lambda }^{[1]}(U_{\lambda
,i})}{g_{\lambda }^{[1]}(U_{\lambda ,i})}\right\}  \\
&=&\frac{1}{\sqrt{n}}\sum_{i=1}^{n}\frac{g_{\psi ,\lambda }^{[1]}(U_{\lambda
,i})}{g_{\lambda }^{[1]}(U_{\lambda ,i})}\left\{ D_{i}\varphi
(W_{i})-g_{\varphi ,\lambda }^{[1]}(U_{\lambda ,i})\right\}  \\
&&+\frac{1}{\sqrt{n}}\sum_{i=1}^{n}\frac{g_{\psi ,\lambda }^{[1]}(U_{\lambda
,i})}{g_{\lambda }^{[1]}(U_{\lambda ,i})}\frac{g_{\varphi ,\lambda
}^{[1]}(U_{\lambda ,i})}{g_{\lambda }^{[1]}(U_{\lambda ,i})}\left\{
g_{\lambda }^{[1]}(U_{\lambda ,i})-D_{i}\right\} +o_{P}(1) \\
&=&\frac{1}{\sqrt{n}}\sum_{i=1}^{n}\frac{g_{\psi ,\lambda }^{[1]}(U_{\lambda
,i})}{g_{\lambda }^{[1]}(U_{\lambda ,i})}D_{i}\left\{ \varphi (W_{i})-\frac{%
g_{\varphi ,\lambda }^{[1]}(U_{\lambda ,i})}{g_{\lambda }^{[1]}(U_{\lambda
,i})}\right\} +o_{P}(1).
\end{eqnarray*}%
\ Note that%
\begin{eqnarray*}
&&\mathbf{E}\left[ \frac{g_{\psi ,\lambda }^{[1]}(U_{\lambda ,i})}{%
g_{\lambda }^{[1]}(U_{\lambda ,i})}D_{i}\left\{ \varphi (W_{i})-\frac{%
g_{\varphi ,\lambda }^{[1]}(U_{\lambda ,i})}{g_{\lambda }^{[1]}(U_{\lambda
,i})}\right\} \right]  \\
&=&\mathbf{E}\left[ \mathbf{E}\left[ g_{\psi ,\lambda }^{[1]}(U_{\lambda
,i})\left\{ \varphi (W_{i})-\frac{g_{\varphi ,\lambda }^{[1]}(U_{\lambda ,i})%
}{g_{\lambda }^{[1]}(U_{\lambda ,i})}\right\} |U_{\lambda ,i},D_{i}=1\right] %
\right]  \\
&=&\mathbf{E}\left[ g_{\psi ,\lambda }^{[1]}(U_{\lambda ,i})\mathbf{E}\left[
\varphi (W_{i})-\frac{g_{\varphi ,\lambda }^{[1]}(U_{\lambda ,i})}{%
g_{\lambda }^{[1]}(U_{\lambda ,i})}|U_{\lambda ,i},D_{i}=1\right] \right] =0.
\end{eqnarray*}%
Therefore, using similar arguments in the proof of Lemma B2, we conclude
that uniformly over $(\lambda ,\varphi ,\psi )\in \Lambda _{n}\times \Phi
\times \Psi ,$%
\begin{equation*}
\nu _{n,1}(\lambda ,\varphi ,\psi )=\frac{1}{P_{1}}\cdot \frac{1}{\sqrt{n}}%
\sum_{i=1}^{n}\frac{g_{\psi ,\lambda }^{[1]}(U_{\lambda ,i})}{g_{\lambda
}^{[1]}(U_{\lambda ,i})}D_{i}\left\{ \varphi (W_{i})-\frac{g_{\varphi
,\lambda }^{[1]}(U_{\lambda ,i})}{g_{\lambda }^{[1]}(U_{\lambda ,i})}%
\right\} +o_{P}(1).
\end{equation*}

\noindent (ii) Write%
\begin{equation*}
\nu _{n,0}(\lambda ,\varphi ,\psi )=\frac{1}{\frac{1}{n}\sum_{i=1}^{n}D_{i}}%
\cdot \frac{1}{\sqrt{n}}\sum_{i=1}^{n}\psi (S_{i})D_{i}\left\{ \frac{\hat{g}%
_{\varphi ,\lambda ,i}^{[0]}(U_{n,\lambda ,i})}{\hat{g}_{\lambda
,i}^{[0]}(U_{n,\lambda ,i})}-\frac{g_{\varphi ,\lambda }^{[0]}(U_{\lambda
,i})}{g_{\lambda }^{[0]}(U_{\lambda ,i})}\right\} .
\end{equation*}%
Using the arguments in the proof of Lemma B2, we can write%
\begin{eqnarray*}
&&\frac{1}{\sqrt{n}}\sum_{i=1}^{n}\psi (S_{i})D_{i}\left\{ \frac{\hat{g}%
_{\varphi ,\lambda ,i}^{[0]}(U_{n,\lambda ,i})}{\hat{g}_{\lambda
,i}^{[0]}(U_{n,\lambda ,i})}-\frac{g_{\varphi ,\lambda }^{[0]}(U_{\lambda
,i})}{g_{\lambda }^{[0]}(U_{\lambda ,i})}\right\}  \\
&=&\frac{1}{\sqrt{n}}\sum_{i=1}^{n}\frac{\psi (S_{i})D_{i}}{g_{\lambda
}^{[0]}(U_{\lambda ,i})}\left\{ \hat{g}_{\varphi ,\lambda
,i}^{[0]}(U_{n,\lambda ,i})-g_{\varphi ,\lambda }^{[0]}(U_{\lambda
,i})\right\}  \\
&&+\frac{1}{\sqrt{n}}\sum_{i=1}^{n}\psi (S_{i})D_{i}\frac{g_{\varphi
,\lambda }^{[0]}(U_{\lambda ,i})}{(g_{\lambda }^{[0]}(U_{\lambda ,i}))^{2}}%
\left\{ g_{\lambda }^{[0]}(U_{\lambda ,i})-\hat{g}_{\lambda
,i}^{[0]}(U_{n,\lambda ,i})\right\} +o_{P}(1) \\
&=&\frac{1}{\sqrt{n}}\sum_{i=1}^{n}\frac{g_{\psi ,\lambda }^{[1]}(U_{\lambda
,i})}{g_{\lambda }^{[0]}(U_{\lambda ,i})}\left\{ (1-D_{i})\varphi
(W_{i})-g_{\varphi ,\lambda }^{[0]}(U_{\lambda ,i})\right\}  \\
&&+\frac{1}{\sqrt{n}}\sum_{i=1}^{n}\frac{g_{\psi ,\lambda }^{[1]}(U_{\lambda
,i})}{g_{\lambda }^{[0]}(U_{\lambda ,i})}\frac{g_{\varphi ,\lambda
}^{[0]}(U_{\lambda ,i})}{g_{\lambda }^{[0]}(U_{\lambda ,i})}\left\{
g_{\lambda }^{[0]}(U_{\lambda ,i})-(1-D_{i})\right\} +o_{P}(1) \\
&=&\frac{1}{\sqrt{n}}\sum_{i=1}^{n}\frac{g_{\psi ,\lambda }^{[1]}(U_{\lambda
,i})}{g_{\lambda }^{[0]}(U_{\lambda ,i})}(1-D_{i})\left\{ \varphi (W_{i})-%
\frac{g_{\varphi ,\lambda }^{[0]}(U_{\lambda ,i})}{g_{\lambda
}^{[0]}(U_{\lambda ,i})}\right\} +o_{P}(1).
\end{eqnarray*}%
Note that $\mathbf{E}\left[ \frac{g_{\psi ,\lambda }^{[1]}(U_{\lambda ,i})}{%
g_{\lambda }^{[0]}(U_{\lambda ,i})}(1-D_{i})\left\{ \varphi (W_{i})-\frac{%
g_{\varphi ,\lambda }^{[0]}(U_{\lambda ,i})}{g_{\lambda }^{[0]}(U_{\lambda
,i})}\right\} \right] $ is equal to%
\begin{eqnarray*}
&&\mathbf{E}\left[ \mathbf{E}\left[ \frac{g_{\psi ,\lambda
}^{[1]}(U_{\lambda ,i})}{g_{\lambda }^{[0]}(U_{\lambda ,i})}\left\{ \varphi
(W_{i})-\frac{g_{\varphi ,\lambda }^{[0]}(U_{\lambda ,i})}{g_{\lambda
}^{[0]}(U_{\lambda ,i})}\right\} |U_{\lambda ,i},D_{i}=0\right] \right]  \\
&=&\mathbf{E}\left[ \frac{g_{\psi ,\lambda }^{[1]}(U_{\lambda ,i})}{%
g_{\lambda }^{[0]}(U_{\lambda ,i})}\mathbf{E}\left[ \varphi (W_{i})-\frac{%
g_{\varphi ,\lambda }^{[0]}(U_{\lambda ,i})}{g_{\lambda }^{[0]}(U_{\lambda
,i})}|U_{\lambda ,i},D_{i}=0\right] \right] =0.
\end{eqnarray*}%
From similar arguments in the proof of Lemma B2, uniformly over $(\lambda
,\varphi ,\psi )\in \Lambda _{n}\times \Phi \times \Psi ,$%
\begin{equation*}
\nu _{n,0}(\lambda ,\varphi ,\psi )=\frac{1}{P_{1}}\cdot \frac{1}{\sqrt{n}}%
\sum_{i=1}^{n}\frac{g_{\psi ,\lambda }^{[1]}(U_{\lambda ,i})}{g_{\lambda
}^{[0]}(U_{\lambda ,i})}(1-D_{i})\left\{ \varphi (W_{i})-\frac{g_{\varphi
,\lambda }^{[0]}(U_{\lambda ,i})}{g_{\lambda }^{[0]}(U_{\lambda ,i})}%
\right\} +o_{P}(1).
\end{equation*}%
$\blacksquare $

\section{Derivation of Asymptotic Covariance Matrices}

\subsection{Example 1: Sample Selection Model with Conditional Median
Restrictions}

Let $u_{Z,i}=Z_{i}-\mathbf{E}\left[ Z_{i}|U_{0,i},D_{i}=1\right] $ and $%
u_{Y,i}=Y_{i}-\mathbf{E}\left[ Y_{i}|U_{0,i},D_{i}=1\right] .$ Let $\tilde{S}%
_{ZZ},$ $\tilde{S}_{ZY},$ $\tilde{\mu}_{Z}(u)$ and $\tilde{\mu}_{Y}(u)$ be $%
\hat{S}_{ZZ},$ $\hat{S}_{ZY},$ $\hat{\mu}_{Z}(u)$ and $\hat{\mu}_{Y}(u)$
except that $\hat{\theta}$ is replaced by $\theta _{0}$. Also, let $\tilde{U}%
_{i}$ be $\hat{U}_{i}$ except that $\hat{\theta}$ is replaced by $\theta
_{0} $. Write $\tilde{S}_{ZZ}-S_{ZZ}=B_{1n}+B_{2n}$, where%
\begin{eqnarray*}
B_{1n} &\equiv &\frac{1}{\sum_{i=1}^{n}D_{i}}\sum_{i=1}^{n}D_{i}\left( 
\tilde{u}_{Z,i}\tilde{u}_{Z,i}^{\top }-u_{Z,i}u_{Z,i}^{\top }\right) \\
B_{2n} &\equiv &\frac{1}{\sum_{i=1}^{n}D_{i}}\sum_{i=1}^{n}D_{i}\left(
u_{Z,i}u_{Z,i}^{\top }-\mathbf{E}[u_{Z,i}u_{Z,i}^{\top }|D_{i}=1]\right) 
\text{,}
\end{eqnarray*}%
and $\tilde{u}_{Z,i}=Z_{i}-\tilde{\mu}_{Z}(\tilde{U}_{i})$ and $\tilde{u}%
_{Y,i}=Y_{i}-\tilde{\mu}_{Y}(\tilde{U}_{i})$. Under regularity conditions,
we can write $B_{1n}$ as%
\begin{equation*}
\frac{1}{\sum_{i=1}^{n}D_{i}}\sum_{i=1}^{n}D_{i}u_{Z,i}\{\tilde{u}%
_{Z,i}-u_{Z,i}\}^{\top }+\frac{1}{\sum_{i=1}^{n}D_{i}}\sum_{i=1}^{n}D_{i}\{%
\tilde{u}_{Z,i}-u_{Z,i}\}u_{Z,i}^{\top }+o_{P}(1/\sqrt{n}).
\end{equation*}%
Using Lemma B3(i), we find that both the sums above are equal to $o_{P}(1/%
\sqrt{n})$. Following similar arguments for $\tilde{S}_{ZY}-S_{ZY}$, we
conclude that 
\begin{eqnarray}
\tilde{S}_{ZZ}-S_{ZZ} &=&\frac{1}{\sum_{i=1}^{n}D_{i}}\sum_{i=1}^{n}D_{i}%
\left( u_{Z,i}u_{Z,i}^{\top }-\mathbf{E}[u_{Z,i}u_{Z,i}^{\top
}|D_{i}=1]\right) +o_{P}(1/\sqrt{n})  \label{state} \\
\tilde{S}_{ZY}-S_{ZY} &=&\frac{1}{\sum_{i=1}^{n}D_{i}}\sum_{i=1}^{n}D_{i}%
\left( u_{Z,i}u_{Y,i}-\mathbf{E}[u_{Z,i}u_{Y,i}|D_{i}=1]\right) +o_{P}(1/%
\sqrt{n}).  \notag
\end{eqnarray}

First write $\tilde{\beta}-\beta _{0}=A_{1n}+A_{2n}$, where%
\begin{equation*}
A_{1n}\equiv \{\tilde{S}_{ZZ}^{-1}-S_{ZZ}^{-1}\}\tilde{S}_{ZY}\text{ and }%
A_{2n}\equiv S_{ZZ}^{-1}\{\tilde{S}_{ZY}-S_{ZY}\}.
\end{equation*}%
From (\ref{state}), $||[\tilde{S}_{ZY}\ \vdots \ \tilde{S}_{ZZ}]-[S_{ZY}\
\vdots \ S_{ZZ}]||=O_{P}(1/\sqrt{n}).$ Hence%
\begin{eqnarray*}
A_{1n} &=&\tilde{S}_{ZZ}^{-1}\{S_{ZZ}-\tilde{S}_{ZZ}\}S_{ZZ}^{-1}\tilde{S}%
_{ZY}=S_{ZZ}^{-1}\{S_{ZZ}-\tilde{S}_{ZZ}\}S_{ZZ}^{-1}\tilde{S}_{ZY}+o_{P}(1/%
\sqrt{n}) \\
&=&S_{ZZ}^{-1}\{S_{ZZ}-\tilde{S}_{ZZ}\}\beta _{0}+o_{P}(1/\sqrt{n}),
\end{eqnarray*}%
because $\beta _{0}=S_{ZZ}^{-1}S_{ZY}$. Therefore, from (\ref{state}), $%
\sqrt{n}\{\tilde{\beta}-\beta _{0}\}$ is equal to%
\begin{eqnarray}
&&-S_{ZZ}^{-1}\left( \frac{1}{\sum_{i=1}^{n}D_{i}}\sum_{i=1}^{n}D_{i}\left(
u_{Z,i}u_{Z,i}^{\top }-\mathbf{E}[u_{Z,i}u_{Z,i}^{\top }|D_{i}=1]\right)
\right) \beta _{0}  \label{dec2} \\
&&+S_{ZZ}^{-1}\left( \frac{1}{\sum_{i=1}^{n}D_{i}}\sum_{i=1}^{n}D_{i}\left(
u_{Z,i}u_{Y,i}-\mathbf{E}[u_{Z,i}u_{Y,i}|D_{i}=1]\right) \right) +o_{P}(1/%
\sqrt{n}).  \notag
\end{eqnarray}%
As for the first term, observe that $u_{Z,i}^{\top }\beta
_{0}=u_{Y,i}-u_{v,i},$ where $u_{v,i}=v_{i}-\mathbf{E}\left[
v_{i}|U_{0,i},D_{i}=1\right] $. From this, we find that%
\begin{eqnarray*}
&&\left( u_{Z,i}u_{Z,i}^{\top }-\mathbf{E}[u_{Z,i}u_{Z,i}^{\top
}|D_{i}=1]\right) \beta _{0} \\
&=&\left( u_{Z,i}u_{Y,i}-\mathbf{E}[u_{Z,i}u_{Y,i}|D_{i}=1]\right) -\left(
u_{Z,i}u_{v,i}-\mathbf{E}[u_{Z,i}u_{v,i}|D_{i}=1]\right) .
\end{eqnarray*}%
Plugging this into the first sum of (\ref{dec2}), we find that $\tilde{\beta}%
-\beta _{0}=S_{ZZ}^{-1}\xi _{n}+o_{P}(1/\sqrt{n}),$ where%
\begin{equation*}
\xi _{n}\equiv \frac{\sqrt{n}}{\sum_{i=1}^{n}D_{i}}\sum_{i=1}^{n}D_{i}\left(
u_{Z,i}u_{v,i}-\mathbf{E}\left[ u_{Z,i}u_{v,i}|D_{i}=1\right] \right) .
\end{equation*}%
Therefore the asymptotic variance is obtained through Assumption SS0 (i). $%
\blacksquare $

\subsection{Example 2: Single-Index Matching Estimators of Treatment Effects
on the Treated}

Define $\tilde{\mu}(\tilde{U}_{i})$ to be $\hat{\mu}(\hat{U}_{i})$ except
that $\hat{\theta}$ is replaced by $\theta _{0}$. Write $\sqrt{n}(\tilde{%
\beta}-\beta _{0})$ as%
\begin{eqnarray*}
&&\frac{1}{\frac{1}{n}\sum_{i=1}^{n}Z_{i}\sqrt{n}}\sum_{i=1}^{n}\left\{
Z_{i}\left( Y_{i}-\mu _{0}(U_{0,i})\right) -\beta _{0}\right\}  \\
&&+\frac{1}{\frac{1}{n}\sum_{i=1}^{n}Z_{i}\sqrt{n}}\sum_{i=1}^{n}Z_{i}\left(
\mu _{0}(U_{0,i})-\tilde{\mu}(\tilde{U}_{i})\right) \equiv A_{1n}+A_{2n}%
\text{, say.}
\end{eqnarray*}%
\newline
Let $P_{d}\equiv P\{Z=d\}$, $d\in \{0,1\}$. As for $A_{1n},$ it is not hard
to see that 
\begin{eqnarray*}
A_{1n} &=&\frac{1}{\frac{1}{n}\sum_{i=1}^{n}Z_{i}\sqrt{n}}%
\sum_{i=1}^{n}Z_{i}\left( Y_{i}-\mu _{1}(U_{0,i})\right) +\frac{1}{\frac{1}{n%
}\sum_{i=1}^{n}Z_{i}\sqrt{n}}\sum_{i=1}^{n}Z_{i}\left\{ \mu
_{1}(U_{0,i})-\mu _{0}(U_{0,i})-\beta _{0}\right\}  \\
&=&\frac{1}{P_{1}\sqrt{n}}\sum_{i=1}^{n}Z_{i}\left( Y_{i}-\mu
_{1}(U_{0,i})\right) +\frac{1}{P_{1}\sqrt{n}}\sum_{i=1}^{n}Z_{i}\left( \mu
_{1}(U_{0,i})-\mu _{0}(U_{0,i})-\beta _{0}\right) +o_{P}(1).
\end{eqnarray*}%
As for $A_{2n}$, we apply Lemma B3(ii) to deduce that it is equal to%
\begin{equation*}
\frac{1}{P_{1}\sqrt{n}}\sum_{i=1}^{n}\frac{(1-Z_{i})P(U_{0,i})}{1-P(U_{0,i})}%
\left( \mu _{0}(U_{0,i})-Y_{i}\right) +o_{P}(1).
\end{equation*}%
Combining $A_{1n}$ and $A_{2n}$, we find that $\sqrt{n}(\tilde{\beta}-\beta
_{0})=\frac{1}{\sqrt{n}}\sum_{i=1}^{n}\gamma _{i}+o_{P}(1)$, where%
\begin{equation*}
\gamma _{i}=\frac{Z_{i}\varepsilon _{1,i}}{P_{1}}-\frac{(1-Z_{i})P(U_{0,i})%
\varepsilon _{0,i}}{(1-P(U_{0,i}))P_{1}}+\frac{1}{P_{1}}Z_{i}\left( \mu
_{1}(U_{0,i})-\mu _{0}(U_{0,i})-\beta _{0}\right) ,
\end{equation*}%
and $\varepsilon _{d,i}=Y_{i}-\mu _{d}(U_{0,i})$, $d\in \{0,1\}$. Hence $%
V_{SM}=\mathbf{E}\gamma _{i}\gamma _{i}^{\top }$.

\section{Acknowledgements}

Part of the result in this paper was circulated in a draft titled
"Bootstrapping Semiparametric Models with Single-Index Nuisance
Parameters."\ I thank Xiaohong Chen, Stefan Hoderlein, Simon Lee, Frank
Schorfheide and seminar participants at the Greater New York Econometrics
Colloquium at Princeton University for valuable comments for the earlier
version of this paper. I also thank Flavio Cunha for providing me with the
data sets that I used for the empirical study, and an associate editor and
two referees for comments that led to a substantial improvement of the
paper. All errors are mine.

\end{document}